# Acoustic Wave Modulation of Gap Plasmon Cavities

Skyler Peitso Selvin[1,2]*, Majid Esfandyarpour[1], Anqi Ji[1,2], Yan Joe Lee[1,3], Colin Yule[1,4], Jung-Hwan Song[1], Mohammad Taghinejad[1], Mark L. Brongersma[1,3]*

[1]Geballe Laboratory for Advanced Materials, Stanford University; Stanford, CA, USA.
[2]Department of Electrical Engineering, Stanford University; Stanford, 94305, USA
[3]Department of Material Science and Engineering, Stanford University; Stanford, 94305, USA
[4]Department of Applied Physics, Stanford University; Stanford, CA 94305, USA
*Corresponding authors. Email: selvin@stanford.edu, brongersma@stanford.edu

**Abstract:**

The role of metallic nanostructures in nanophotonics is expected to expand if ways to electrically manipulate their optical resonances at high speed can be identified. Here, we capitalize on electrically-driven surface acoustic waves and the extreme light concentration afforded by gap plasmons to achieve this goal. We place gold nanoparticles in a particle-on-mirror configuration with a few-nanometer-thick, compressible polymer spacer. Surface acoustic waves are then used to tune light scattering at speeds approaching the GHz regime. We observe evidence that the surface acoustic waves produce mechanical deformations in the polymer, and that ensuing nonlinear mechanical dynamics lead to unexpectedly large levels of strain and spectral tuning. Our approach provides a design strategy for electrically driven dynamic metasurfaces and fundamental explorations of high-frequency, polymer dynamics in ultra-confined geometries.

The ability of metasurfaces to actively control the flow of light at the nanoscale has had a transformative impact on a wide variety of optical technologies and enabled entirely new functionalities (*1-3*). Dynamic tuning of these thin optical elements requires strong and highly-tunable light-matter interaction. This can be achieved by employing resonant nanostructures and modulating their permittivity or morphology. Various physical principles have been explored to achieve such tuning, including structural phase transformations (*4–6*), thermo-optic effects (*7*), electrochemistry (*8–10*), liquid crystals (*11, 12*), carrier injection (*13–16*), and mechanical motion (*17–25*). However, it remains challenging to realize metasurface building blocks that deliver large, high-speed changes in their scattering properties. It is therefore critical to continue the search for new mechanisms to dynamically control light scattering from nanostructures.

The Kramers–Kronig relations provide valuable insight into the limitations of refractive index tuning. They dictate that practical changes in the real part of the index are accompanied by undesirable changes in the imaginary part, causing absorption. To skirt this fundamental limitation, we can instead mechanically reconfigure the nanostructures, effectively replacing the material at a specific location with a different one. The achievable speeds for such displacements are constrained by Newton's second law, which states that the force required to mechanically oscillate a structure is proportional to the required displacement multiplied by the frequency squared. As a result, state-of-the-art optical microelectromechanical systems (MEMS) are typically limited to low MHz regimes (*17–25*). To address this challenge, we combine the use of electrically-driven surface acoustic wave (SAW) displacements with the extreme sensitivity of the metal-nanoparticle-on-mirror system to deliver tuning at frequencies approaching 1 GHz.

Metallic nanoparticles efficiently scatter light because they support strong plasmonic resonances. When two metal nanoparticles are brought together, their plasmonic modes hybridize and the optical fields redistribute into the gap. For nanometer-sized gaps, this leads to extreme light concentration and the scattering becomes sensitive to angstrom-level changes in gap size (*26–28*). Similar physics occurs when a metallic nanoparticle is placed above a metallic film in a nanoparticle-on-mirror (NPoM) configuration, where the nanoparticle's electric currents can interact with their mirror images (*29–32*). Some attempts have been made to dynamically tune plasmonic gap resonances using electrochemistry (*33*), carrier-injection (*15, 16*), MEMS (*17, 18*), and thermal expansion (*34–36*), but it remains challenging to achieve high-speed, robust, electrical tuning. We address this by capitalizing on the fact that the size of these gaps are comparable to acoustic displacements and find that SAWs produce observable changes in the optical resonances of an NPoM when a soft material such as polydimethylsiloxane (PDMS) fills the gap. This structure minimizes the mechanical displacement required to effectively tune optical resonances. Polymer materials such as PDMS are chosen because they have low stiffness and can sustain large, fast strains in thin films (*37, 38*).

## Device design and simulation

In our device (Fig. 1A) an interdigitated transducer (IDT) on a piezoelectric lithium niobate (LiNbO₃) substrate launches SAWs to a region where 100-nm-diameter gold nanoparticles (AuNPs) are drop-cast on a PDMS-coated Au film (see (*39*) and fig. S1 for details). The SAWs produce oscillatory vertical and horizontal surface displacements, moving the NPoMs and likely deforming the PDMS in the gap (see Fig. 1B). We deposit PDMS layers with various thicknesses from 2 – 10 nm, creating NPoM systems with light-scattering sensitive to gap size. We monitor these scattering changes in a dark-field optical microscope under white-light illumination. Figure 1C shows an image of an area with several NPoMs on a ~ 4-nm-thick PDMS layer that notably change color when a SAW is activated.

The scattering changes may largely be understood by analyzing the plasmon mode's dependence on gap size, although other mechanisms (e.g. changes in index of the gap filler) may also contribute. Under s-polarized dark-field illumination, our faceted AuNPs support two gap plasmon modes in the visible and near-infrared (NIR) spectral regions (*32*). Figure 1D maps the dependence of the scattering efficiency on gap size and illumination wavelength (also see figs. S2-4). This efficiency is defined as the ratio of the scattering cross section $\sigma_{sc}$ to the NP's geometric area $\sigma_{NP}$. The two modes are the fundamental $s_{11}$ and a higher-order $s_{12}$ mode that result from resonating plasmons inside the gap (Fig.1E) (*31*). The cavity length for these is set approximately by the size of the bottom facet of the AuNP, which we model as circular with a 40 nm diameter. Mode $s_{11}$ shows the strongest scattering dependence on gap size. Cuts from panel 1D show that an increase in gap size from 4 to 6 nm comes with a notable increase in the scattering intensity as well as a blue shift in the resonance wavelength $\lambda_{11}$ that well-exceeds the linewidth (Fig. 1F). Both these changes contribute to the color modification. When the SAW is off, the color appears greenish due to the $s_{12}$ mode around $\lambda_{12} \approx 550$ nm. Upon SAW activation, the mode $s_{11}$ undergoes a blueshift from the invisible NIR ($\lambda_{11} \approx 670$ nm) to the visible red ($\lambda_{11} \approx 620$ nm) and becomes appreciably brighter, causing the color change from greenish to reddish.

The details of the SAW generation and propagation are shown in Fig. 2, where we place a chip with an IDT under an optical microscope and apply radio-frequency (RF) signals. The IDT generates spatially periodic electric fields across the surface of the LiNbO₃. Through the piezoelectric effect, these induce periodic mechanical displacements that propagate across the surface as SAWs. Measurements of light diffraction from the optical phase grating produced by the SAW indicate that the IDT produces ~0.3 nm vertical surface displacements across 200–700 MHz (Fig. 2C) (see (*39*), fig. S5 and (*40–42*)). This also agrees with our acoustic simulations.

**Dynamic optical response**

When the SAW is activated, we observe changes in optical scattering from NPoMs that are exposed to the SAW. Figure 3 shows the changes in the scattering spectrum of one representative NPoM when a 670 MHz SAW is activated with an RF power of 2 W into the device. We monitor the dynamic scattering using a time-correlated single-photon counting system and a tunable laser (details in (*39*) and figs. S6–S8). Before the SAW is active, the NPoM displays plasmon resonances at $\lambda_{11} \approx 680\ nm$ and $\lambda_{12} \approx 540\ nm$. When the SAW is activated, both resonances blueshift during the initial few microseconds and $\lambda_{11}$ shifts to 620 nm (Fig. 3A, 3B). While the SAW is active, we also see faster scattering changes at the SAW frequency (Fig. 3C) where $\lambda_{11}$ oscillates approximately sinusoidally between 622 and 615 nm. We term these fast scattering changes the fast optical response (FOR), and we dub the slower change that occurs when the SAW is first activated the slow optical response (SOR). Both provide substantial optical modulations. At a wavelength of 580 nm, the FOR provides a ~ 25% scattering modulation. The SOR offers even higher modulation of ~ 85% at 620 nm (see figs. S7–S10).

To appreciate the magnitude of the scattering changes caused by the SOR, we record movies of our devices. Movies S1 and S2 show dark-field recordings while we turn the SAW on and off at 1 Hz. Here, most of the NPoMs initially have $\lambda_{11}$ in the NIR. The NPoMs exposed to the SAW become brighter and redder in color as $\lambda_{11}$ shifts from the NIR into the visible. Further, we see that in areas where the SAW does not reach—for example, on the electrical pad in Movie

S1, or the area outside of the SAW beam in Movie S2—the NPoMs' color remains constant, indicating that the SAW is driving the scattering modulation.

**Proposed mechanism for the SAW-induced color tuning**
The optical modulation may be caused by a variety of physical effects. The SAW produces surface displacements, but also generates heat and electric fields (*42*). Heat is an unlikely cause for the spectral shifts given the temperature stability of the mechanical and optical properties of PDMS (see (*39*) and figs. S11-15). Furthermore, the electric fields associated with the SAW are also negligible (fig. S16). Given the difficulty of performing high-speed, *in situ* optical and structural measurements inside the nanoscale gap of the NPoMs, it is challenging to obtain direct evidence for possible changes in the local morphology and/or optical properties. However, although other physical effects cannot be fully ruled out, in the following we argue that mechanical deformations of the PDMS layer are a likely cause, and we provide a mechanical model that is in good agreement with our optical observations.

 To approximate the gap changes that the SAW may induce, we initially make simplifying assumptions that the shifts in the scattering spectra can be fully ascribed to changes in the gap size and that the polymer maintains the shape of a film. By comparing the plasmonic simulations (Fig. 1D) and our experimental plasmonic shifts (Fig. 3A–E), we can estimate the gap size from the location of the measured plasmonic resonance $\lambda_{11}$. In making this estimate, it is important to note that the scattering spectrum of the NPoMs sensitively depends on the facet size and shape (*32*) which will vary from particle to particle and are difficult to determine. However, for this NPoM we find that choosing a 40 nm diameter circular facet allows the simulations to closely match the experimental spectra for gap sizes close to the polymer thickness measured with ellipsometry (2–4 nm). Shown on the left *y*-axis of Fig. 3F, the SOR is consistent with a slow gap size increase from approximately 3.6 to 6.4 nm, and the FOR is consistent with a near-harmonic gap oscillation with an amplitude of about 0.3 nm. We name the slowly changing gap size that may explain SOR the slow mechanical response (SMR), and the quickly varying gap size that may explain the FOR the fast mechanical response (FMR).

 Whereas PDMS is known to conserve its refractive index during such large strains (*43*), it's possible the polymer conformation inside the gap changes (e.g. necking), which can contribute to the observed blueshift (fig. S3B). Conformational changes, however, are difficult to assess at the nm and ns scales. Therefore, we assume that the polymer remains uniformly distributed under the particle so that it can be modeled as a dielectric slab.

**Mechanical model for the NPoM dynamics**
The dynamics of both the FMR and SMR can be captured with a lumped spring-mass-damper model as shown in Fig. 4A, where the AuNP is the mass and the elastomer is modeled as a spring and dashpot in parallel. This material model, known as the Voigt model (*44*), effectively places the AuNP in a potential well. An elastic force $F_{spr} = -k(g - g_0)$ captures the elastic behavior of the polymer, and a drag force $F_{vis} = -\zeta \dot{g}$ describes its viscous behavior. Here, $g$ is the gap size, $\dot{g}$ is its rate of change, $k$ is the spring constant, $\zeta$ is the drag constant, and $g_0$ is the initial gap thickness at rest. The Voigt element is driven with a sinusoidal displacement of amplitude $b_0$ that represents the vertical surface displacement of the SAW.

 Since the SMR does not move at the SAW frequency, it cannot be described by a linear model with constant $\zeta$ and $k$. Indeed, as shown in Figs. 4B and 4C, we would expect an asymmetrical structure such as ours to have asymmetrical spring and viscous forces. At smaller gaps, the lateral displacement of polymer material is more restricted due to the Au surfaces at its top and bottom. When the AuNP is pushed down into the polymer layer it must displace all the polymer molecules directly below it. However, when it is pulled upward, it only pulls the subset of those molecules to which it is chemically attached. Additionally, van der Waals and other surface forces become stronger at smaller gaps. Therefore, both $k$ and $\zeta$ should be asymmetrical and larger at smaller gaps (see fig. S17) (45–48).

When we introduce an asymmetrical $k$, it generates an asymmetrical potential. However, its minimum remains at $g = g_0$, and every cycle the AuNP must travel through this point. It is rather the asymmetry in $\zeta$ that can explain the SMR, where the gap slowly increases over many acoustic cycles and does not return to $g_0$ while the SAW is on. When the drag coefficient is smaller for upward motion, we find that the AuNP experiences a net slow force in the direction of lower drag, sometimes called the vibrational force (46).

For the simplest model to demonstrate this concept, we use a linear spring and an asymmetrical dashpot that has unequal drag coefficients in the downward and upward directions:

$$F_{vis}(\dot{g}) = -\zeta_{\pm}\dot{g} = \begin{cases} -\zeta_{+}\dot{g} & for \quad \dot{g} \geq 0 \\ -\zeta_{-}\dot{g} & for \quad \dot{g} < 0 \end{cases} \qquad \text{Eq. 1}$$

where $\zeta_{+}$ is the smaller drag coefficient for increasing gap size (AuNP moving away from the substrate), and $\zeta_{-}$ is the larger drag coefficient for decreasing gap size (AuNP moving towards the substrate). This asymmetrical drag force is shown in Fig. 4B. We can expand the gap size as $g(t) = G(t) + \tilde{g}(\omega t)$ to independently analyze the slow ($G$) and fast ($\tilde{g}$) gap changes that govern the SMR and FMR respectively. From the experiments, we find that the fast component is accurately represented by a single harmonic at the drive frequency: $\tilde{g} \approx \tilde{g}_0 \sin(\omega t + \theta)$. When the SAW is turned on and $\tilde{g}_0$ is roughly equal to the driving amplitude $b_0$ (as seen in our experiment Fig. 3F), we can approximately solve the equation of motion for the slow gap size $G$ analytically (Eq. S15). We find that there is a component of the slow force $\langle F_{vis} \rangle$ that is exerted upward (positive force) on the AuNP and that originates from the asymmetric drag:

$$\langle F_{vis} \rangle \approx \frac{\Delta\zeta b_0 \omega}{\pi} - \bar{\zeta}\dot{G} \ . \qquad \text{Eq. 2}$$

Where $\Delta\zeta = \zeta_{-} - \zeta_{+} > 0$ is the asymmetry in the dashpot and $\bar{\zeta} = \frac{\zeta_{-}+\zeta_{+}}{2}$ is its average value. The asymmetry causes every acoustic cycle to push the AuNP up a bit more than it is pulled down. Over many cycles, this results in a slowly acting force and causes the gap size to slowly increase until this vibrational force pointing upward is equal to the spring force pulling downward. The particle then reaches a new equilibrium point, yielding a total change in the slow gap size

$$\Delta G \approx \frac{\Delta\zeta b_0 \omega}{k\pi} \ . \qquad \text{Eq. 3}$$

The full mechanical response of this linear spring-asymmetrical dashpot model to a pulsed SAW is shown in Fig. 4D. Both the approximate analytical solution for $G$ and the numerical solution for $g$ are plotted for a SAW with amplitude $b_0 = 0.36$ nm, $\zeta_{-} = 1.05 \times 10^{-8}$ kg/s, $\zeta_{+} = 1 \times 10^{-8}$ kg/s, $k = 0.1$ N/m, and frequency 670 MHz. The gap size initially starts at $g_0$ when the SAW is off. Then, when the SAW is activated, there is a gap increase that eventually plateaus at a slow gap of $\Delta G + g_0$. Comparing these dynamics to the experimental dynamics in Fig. 3F, we can see that this simple model matches the measurements well.

It is important to note that the SMR requires repeated compressions and expansions of the gap to reveal the asymmetry in the upward and downward drag. At low frequencies, the motion of the Au nanoparticle follows that of the substrate and the gap does not change. Consequently, the SMR will not appear at low frequencies. Only near the mechanical resonance frequency will the phase of the particle's oscillatory motion deviate from that of the substrate. Therefore, we expect an onset in the SOR with increasing frequency, and our experiments indeed show such an onset around 400 MHz (Fig. 4E, green curve). This suggests that the resonance frequencies are ∼ 400 MHz, in good agreement with our theoretical analysis (see (*39*) section S10). Additionally, in agreement with Eq. 3, increasing SAW amplitude provides greater optical modulation (Fig. 4E, blue curve).

Whereas this linear spring-asymmetrical dashpot model nicely captures the SMR, introducing nonlinearity into the spring as well helps understand observed modulations over a range of SAW frequencies and amplitudes (see (*39*) section S11 and figs. S18-S20).

## Concluding Remarks

Mechanical methods to manipulate gap plasmons have been demonstrated before (*17, 18, 35*), but these methods don't deliver robust electrical modulation, high optical efficiency, and high (GHz) speeds as we show here. With acoustically modulated gap plasmons, the interaction of acoustic waves and optical waves is no longer limited by tiny photoelastic changes in refractive index, as is the case of traditional acousto-optic modulators, but is instead controlled by the plasmonic modal properties. Typically, millimeters of length are required to modulate light with acoustics (*49*), but our device notably allows for more than 80% optical modulation with interaction lengths less than 100 nm. Furthermore, unlike typical MEMS that use brittle materials, we observe material strains approaching 100%, eliminating the need for large beams or springs that increase the size and required spacing of the modulators. Indeed, we can place our tunable nanostructures closer than the optical wavelength, allowing the realization of dynamic metasurfaces. Shaping the acoustic excitation is also possible; dynamic acoustic images can shape the optical phase and amplitude over the surface, allowing the creation of dynamic optical holograms. We have already observed that acoustic waves can generate a spatially varying optical scattering over the surface, effectively creating an acoustically modulated optical diffraction grating (see figs. S21, S22, and Movie S3).

From the perspective of polymer physics, the polymer layer likely experiences a strain of 5% at the nanosecond scale and over 50% at the microsecond scale. We expect these strains will increase with higher-frequency devices, which can reach ∼40 GHz (*50*). Very soft and thin polymer layers are difficult to mechanically characterize at such high strains and speeds, and our device enables these measurements over a broad range of high frequencies (*37, 51, 52*).

Finally, the often counterintuitive effects of vibrational mechanics have led to many industrial and scientific techniques at the macroscale, solving problems where an intuitive solution is not apparent (*46*). We hope that these results lend similar inspiration to nanoscale techniques, expand the basis of intuition for nanoengineers, and help develop solutions where conventional methods yield none.

**Acknowledgements:**

The authors would like to acknowledge Eric Appel, Andy Spackowitz, Amir Safavi-Naeini, and Zhenan Bao for valuable discussions. ChatGPT was used to proofread the language of the manuscript, debug code, and translate pseudocode into source code for data analysis and modeling [further details areprovided in the supplementary materials (39)]. Part of this work was performed at the Stanford Nanofabrication Facility (SNF) and Stanford Nano Shared Facilities (SNSF), supported by the National Science Foundation under award ECCS-2026822.

**Funding:**

We acknowledge support from the Department of Energy Grant DE-FG07-ER46426 and from Meta Platforms Incorporated.

**Author Contributions:**
Conceptualization: SPS, ME, MLB
Methodology: SPS
Investigations: SPS, ME, YJL
Visualization: SPS, MT, MLB
Funding acquisition: SPS, MLB
Project administration: SPS, MLB
Supervision: AJ, JHS, MT, MLB
Validation: SPS, ME, AJ, CY, MT, JHS, MLB
Writing – original draft: SPS, MLB
Writing – review and editing: SPS, ME, AJ, CY, MT, JHS, MLB, YJL

SPS, ME, and MLB conceived the concepts of the paper. SPS, ME, and YJL conducted feasibility studies. SPS designed, fabricated, simulated, and tested all devices, and developed the mechanical and thermal theories. CY, JHS, MT, and MLB assisted SPS in verifying and debugging. AJ, JHS, MT, and MLB supervised and provided guidance. SPS and MLB prepared the original manuscript draft with input from all authors during review and editing.

**Competing interests:**
The authors have a patent application that is related to the content in this paper. The details of this patent are below.

    Patent title: Acoustically modulated plasmonic optical resonators.
    Application number: 18144090
    Patent number: US20230359072A1
    Consignee: Leland Stanford Junior University
    Authors: Skyler Selvin, Mark L. Brongersma, Majid Esfandyarpour, Jung-Hwan Song.

**Data and materials availability:**
All data are available in the manuscript, the supplementary material, or deposited at Dryad (*53*).

**List of Supplementary Materials:**
Materials and methods
Supplementary Text

Figs. S1 to S22
Moves S1 to S3
References (*54-72*)

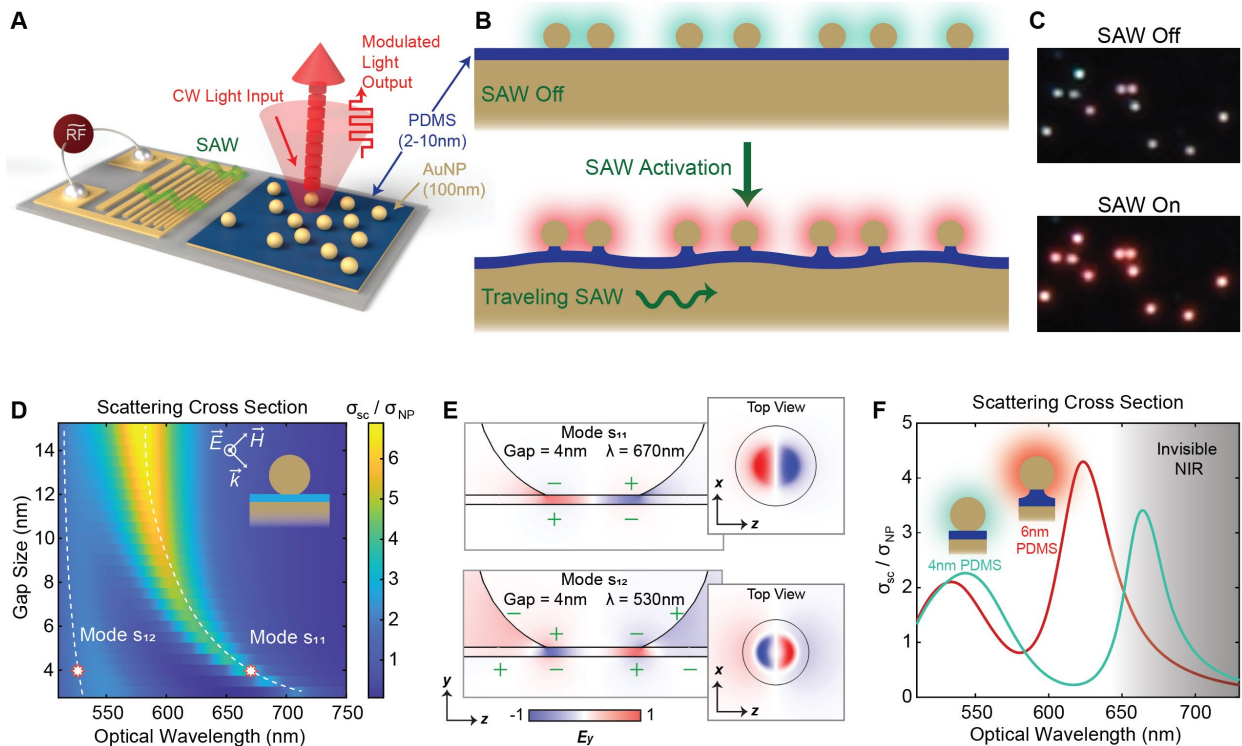

**Fig. 1. Acoustic modulation of Au nanoparticles on an Au mirror.** (**A**) Schematic of the proposed surface acoustic wave (SAW) device capable of acoustically modulating gap plasmons. A chirped interdigitated electrode transducer (IDT) on a LiNbO$_3$ substrate (gray) launches SAWs that can modulate the thickness of a polydimethylsiloxane layer (PDMS, blue) placed in the gap of an Au nanoparticle-on-mirror (NPoM system). The resulting changes in the plasmon resonance can be monitored in a dark-field microscope. (**B**) Drawings of the states where the SAW is inactive (top) and active (bottom). The gap size of the NPoM is expected to change when the SAW is launched. (**C**) Optical dark-field images show clear color changes of the NPoMs in response to the SAW. (**D**) Electromagnetic simulations of a 100-nm-diameter NPoM showing the change in scattering cross-section spectra for various gap sizes. Calculations are performed for s-polarized illumination of an NPoM with a 40 nm diameter facet. (**E**) Vertical amplitude of the scattered electric field $\vec{E}_y$ for each of the two observed gap plasmon modes with a 4 nm gap size. The $s_{11}$ field profile is taken at 670 nm and the $s_{12}$ field is at 530 nm, as indicated by the red stars in (D). (**F**) Spectra of the scattering cross section taken from panel (D) for selected PDMS thicknesses. An increase of the gap from ~4 to ~6 nm is consistent with the experimentally observed color changes of the NPoMs depicted in panel (C).

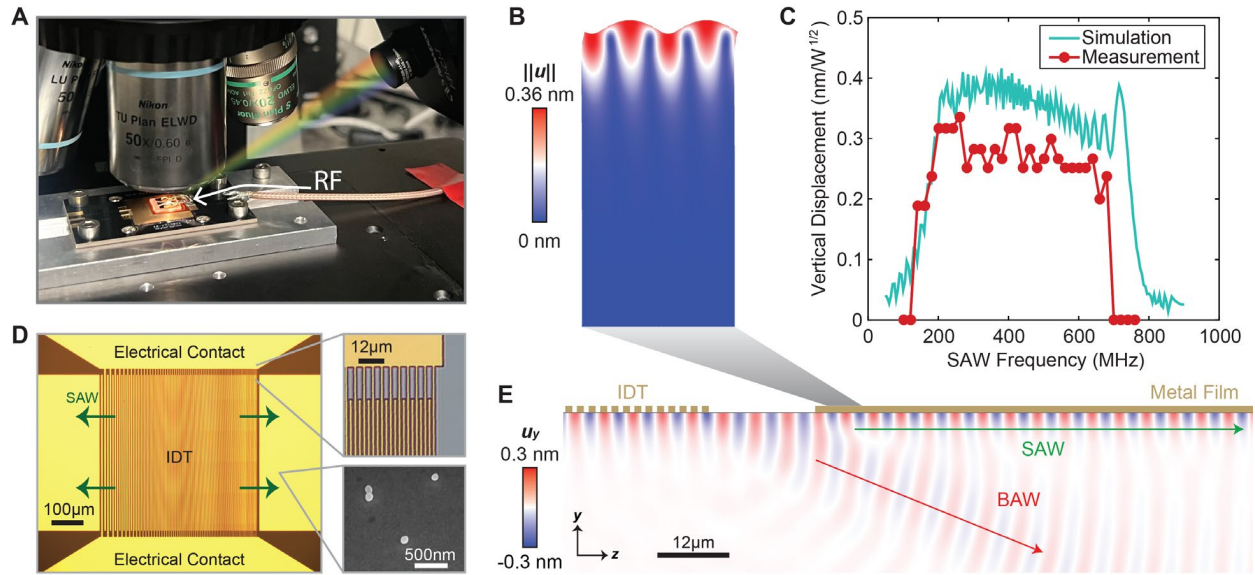

**Fig. 2. Surface acoustic wave transduction**. (**A**) Photograph of the experimental setup, showing the device under a microscope mounted on a printed circuit board. Two watts of RF energy is coupled in via a coaxial cable. The rainbow-colored path illustrates the s-polarized dark-field illumination from the optics on the right. (**B**) Simulated mode shape of the SAW from an electromechanical simulation, showing the total displacement distribution through the thickness of the substrate. The color bar represents the magnitude of the total displacement. (**C**) Vertical displacement of the SAW under the NPoM structures on the Au film, with both simulated and representative measured results shown. (**D**) Optical image of the device, with insets showing an optical image of the IDT fingers and a micrograph of the NPoM atop the metal film to the right of the IDT. (**E**) SAW simulation at 600 MHz showing the vertical displacement amplitude $u_y$ through the thickness of the LiNbO$_3$ substrate. A SAW is observed as a guided wave propagating along the surface of the substrate, while a bulk acoustic wave (BAW) is a parasitic mode emitted from the IDT into the bulk of the substrate.

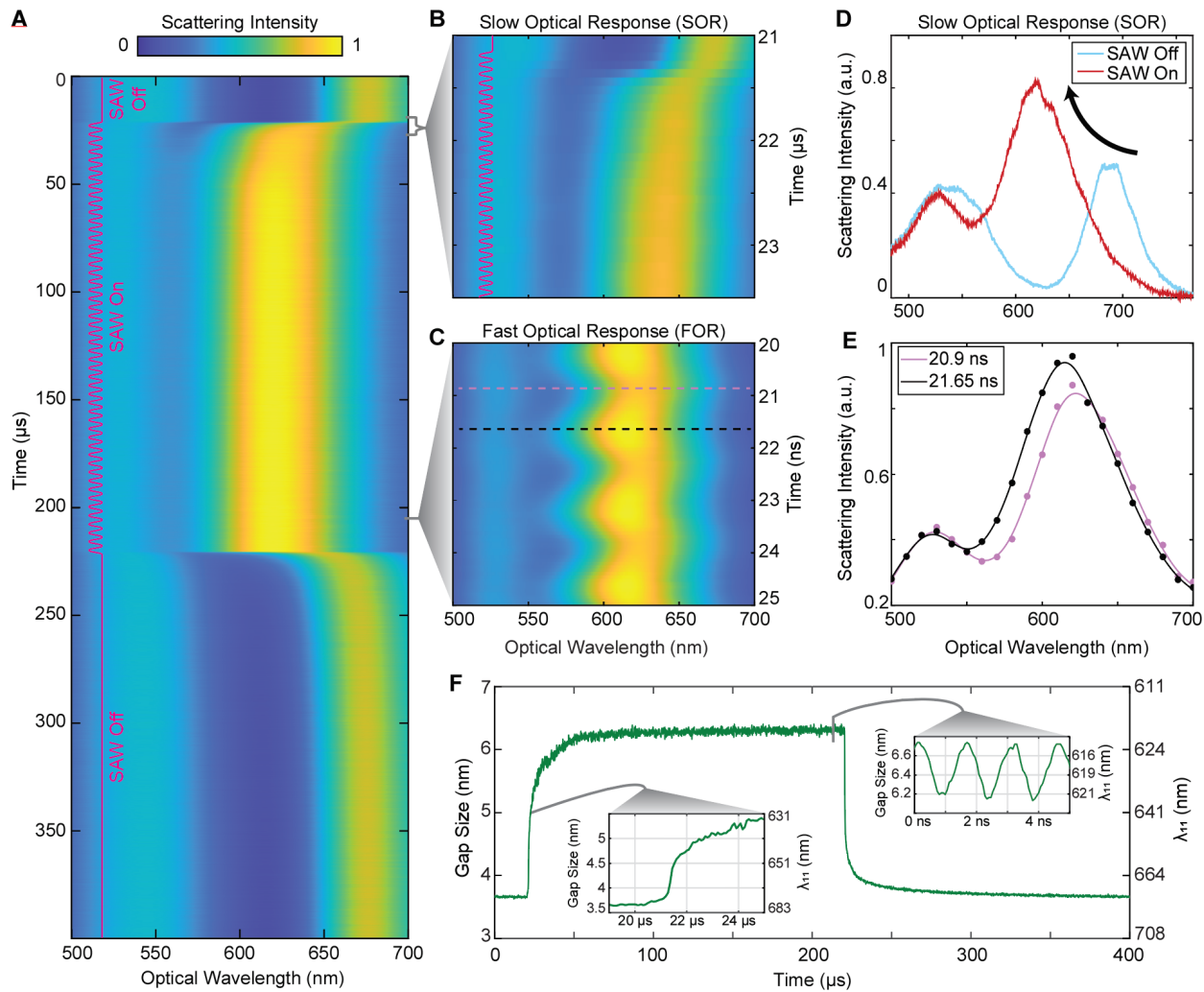

**Fig. 3. Dynamics of SAW-induced optical modulation**. (**A**) Optical scattering spectrum from a single NPoM as we turn the SAW on for a time-period of 200 μs and then turn it off for a period of 200 μs. The red line shows when the SAW is active. The temporal resolution during this measurement is much less than the SAW period and the high frequency fast optical response is averaged. (**B**) Zoomed in optical scattering spectrum of the blue shift that occurs when the SAW is first activated. (**C**) Optical scattering spectrum taken with a temporal resolution exceeding that of the SAW period, thus capturing the fast optical response. We observe small spectral undulations at the SAW frequency. (**D**) Time averaged spectrum of the NPoM when the SAW is active and when it is inactive. A distinct blue shift is apparent when the SAW is active. (**E**) Two spectra taken from panel (C) that show the high-frequency spectral undulations. Each curve is taken at times that are half a SAW temporal period apart (dashed lines on panel C). (**F**) The extrapolated gap size and the peak scattering wavelength over the period of the pulsed SAW. Inlays show the rising response of the gap size, and the gap size as it is changing at high speed with the SAW period.

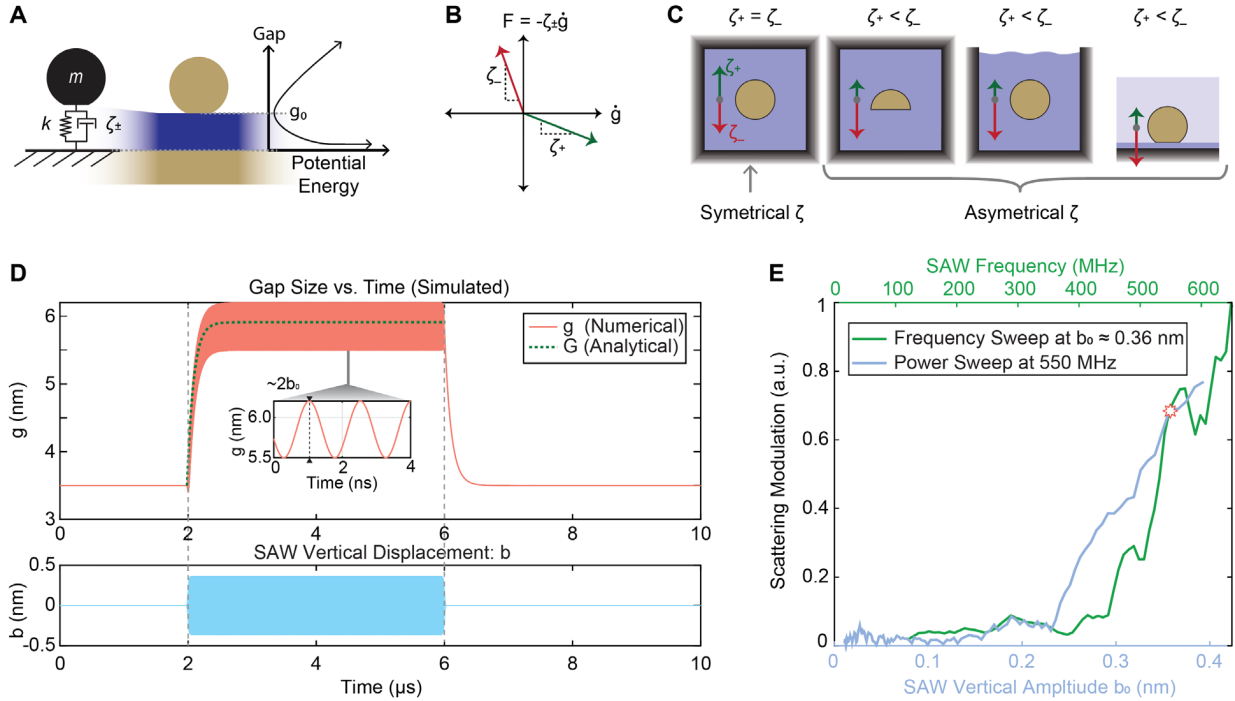

**Fig. 4. Mechanical model**. (**A**) Schematic showing how the NPoM may be modeled as a spring-mass-damper system, with a spring of spring constant $k$ and a damper with an asymmetrical drag coefficient $\zeta_{\pm}$. A schematic of the relevant potential energy versus gap size is also shown. (**B**) Force on the asymmetric dashpot as a function of the gap size rate of change $\dot{g}$. (**C**) Sketch of four different situations where a solid mass (gold domain) is sitting in a viscous fluid (blue domain) with walls at the edge of fluid (black lines), or a free boundary (shown as a wavy ripple in the fluid). The resulting drag coefficient in each of the four examples is shown for the positive (up) and negative (down) directions. On the far left, the geometry is symmetric and thus $\zeta_{+} = \zeta_{-}$. For each of the other three examples, the geometry is asymmetric leading to $\zeta_{+} < \zeta_{-}$. (**D**) Numerical solution to the full equation of motion of mass-spring-asymmetrical damper system shown in (A) using Eq. S7, and analytic solution to the approximate linear equation of motion for the slow gap (Eq. S15). The bottom panel shows the timing of the SAW, and the top panel shows the change in gap size as a result. (**E**) The measured average scattering modulation of many NPoMs detected by red-filtered CCD pixels. The bottom axis and blue curve show a sweep of the SAW amplitude (assuming $0.25\ nm/\sqrt{W}$) at a constant frequency, while the top axis and green curve show a sweep of the SAW frequency for constant amplitude. We see that sweeps of either the amplitude or frequency generate similar trends. The red star indicates the location that both sweeps are 550 MHz and $b_0 \approx 0.36$ nm.

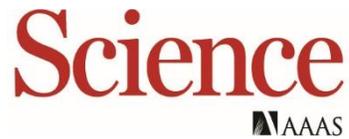

# Supplementary Materials for

## Acoustic wave modulation of gap plasmon cavities

Skyler P. Selvin *et al*.

Corresponding authors: Skyler P. Selvin, selvin@stanford.edu; Mark L. Brongersma, brongersma@stanford.edu

**The PDF file includes:**

Materials and Methods
Supplementary Text
Figs. S1 to S22
References

**Other Supplementary Material for this manuscript includes the following:**

Movies S1 to S3

**Materials and Methods**

<u>SAW measurement and modeling.</u>

In the main text, we describe how a surface acoustic wave (SAW) can drive the mechanical oscillation of a nanoparticle on a mirror (NPoM). In order to build and verify our proposed physical models for the mechanical behavior of the driven NPoMs, it is important to measure and simulate the amplitude of the SAW displacements of the mirror surface in the out-of-plane direction (i.e. the vertical direction). To this end, we use the fact that a SAW turns a metallic surface into a phase grating *(54, 55)*. The amplitude of the phase grating determines how effectively an incident light beam is redirected into various diffracted orders.

Fig. S5 shows our home-built optical setup in which a laser beam illuminates a metal pad during excitation with a SAW. The path length for light that reflects from a metal region where the displacement is positive and the surface is elevated upward is different from the path length for light that strikes a section of the surface that is depressed downward (negative displacement). This path length difference creates a spatially varying, sinusoidal phase profile and the SAW thus acts as a phase grating with a phase amplitude that is dependent on the vertical displacement of the SAW. The power in the reflected $0^{th}$ order beam is linked to the induced phase profile $\phi = \phi_0 \sin(k_{saw}z)$ along the $z$-direction (i.e. along the metal surface) with a characteristic wavenumber $k_{saw}$ is:

$$P_{0th} = P_i[J_0(\phi_0)]^2. \qquad \text{Eq. S1}$$

Here, $P_i$ is the optical power of the incident beam that illuminates the grating, $P_{0th}$ is the optical power in the reflected $0^{th}$-order diffracted beam, $J_0$ is the Bessel function of the $1^{st}$ kind and $0^{th}$-order, and $\phi_0$ is the phase contrast (the amplitude of the sinusoid in the sinusoidal phase grating). The phase contrast for a SAW vertical displacement amplitude of $b_0$ and an incident angle of 45 degrees is (see Fig. S5B):

$$\phi_0 = \frac{2\pi}{\lambda_{HeNe}} b_0 \sqrt{2}. \qquad \text{Eq. S2}$$

Thus, the power in the $0^{th}$-order can be related to the SAW's vertical amplitude as:

$$P_{0th} = P_i \left[ J_0 \left( \frac{2\pi}{\lambda_{HeNe}} b_0 \sqrt{2} \right) \right]^2. \qquad \text{Eq. S3}$$

Using the setup in Fig. S5A, we can measure the power lost in the $0^{th}$-order due to the SAW vertical displacement profile.

The analysis presented above is correct for the traveling SAWs in our experiments (not a standing SAW), which is approximately true for our system where bulk acoustic waves (BAWs) play a minor role (except for the special case as seen in S8).

To numerically model the SAW displacement, we create a two-dimensional piezoelectric finite-element (COMSOL) simulation where the entire interdigitated electrode (IDT), LiNbO₃ substrate, and metal planes are simulated. Perfectly matched layers (PMLs) are placed around the simulation domain to eliminate SAW reflections from the domain edges, creating the same traveling SAWs as we analyze in our experiments. The piezoelectric and mechanical properties of LiNbO₃ and Au are provided by COMSOL, and no mechanical losses are used in the model. The RF input is modeled as a terminal with 1 W of incident power and a 50 Ω intrinsic impedance. The electromagnetic properties of Au are modeled with the complex permittivity $\epsilon = \epsilon_0 - i\,\sigma/\omega_{SAW}$, where $\epsilon_0$ is the permittivity of free space, and $\sigma = 45.6$ MS/m is the electric conductivity at the SAW frequency $\omega_{SAW}$. Both the simulation and experimental results for the SAW amplitude of our device are shown in Fig. 2C of the main text.

The discrepancy between the experimental and simulated results can most likely be attributed to one or more of the following factors:

1. The simulation domain does not include mechanical losses.
2. Fabrication results in more than 50% metallization of the IDT, which decreases its SAW transduction efficiency. This is especially true for the small finger pitch (high SAW frequency) parts of the IDT.
3. The HeNe beam creates a spot size of about 80 μm on the metal plane of the SAW device, and is placed within 500 μm from the IDT. The SAW is attenuated slightly over the length of the spot size as well as over the length of the metal film en route to the HeNe spot.

4. Possible interference of the traveling wave with bulk waves and reflected SAWs.
5. Additional capacitance between the metal electrical contact pads and the ground plane behind the LiNbO$_3$ chip.

Setup to acquire time-resolved spectra of the light scattering from NPoMs.

To measure the time-resolved light scattering spectra of NPoMs modulated by a SAW, we use the experimental setup shown in Fig. S6. The sample is illuminated with s-polarized single-wavelength light from a wavelength-tunable laser while the SAW is chopped for the slow optical response (SOR) and continuously on for the fast optical response (FOR). We change the illumination wavelength and repeat the process across the relevant spectral range from 500 to 700 nm using 10 nm intervals. These time traces are then normalized using lamp illumination (see details below), stitched together, and smoothed in both time and wavelength to produce the spectra presented in Fig. 3 and Fig. S10.

For the single wavelength illumination, we use a supercontinuum laser (NKT Photonics SuperK EXTREME) with an acousto-optic tunable filter (AOTF, NKT Photonics SuperK SELECT) that allows us to vary the illumination wavelength continuously across the spectral range from 500 to 700 nm with a linewidth of about 5 nm. The light coming from the AOTF is fiber-coupled and delivered through an S-polarizer, and then focused onto the sample outside of the 0.6 numerical aperture (NA) of a 50X collection objective. The light that is scattered by the NPoM is then collected by this objective. Using a scanning confocal head, we exclusively select the light coming from an NPoM of interest and couple it into a multimode fiber that is connected to a single photon avalanche detector (SPAD) (Micro Photon Devices PDM $PD-050-CTD-FC). For the broadband illumination, we use a multimode fiber-coupled lamp in place of the single wavelength supercontinuum laser. See Fig. S6 for a diagram of the experimental setup.

The SPAD features a time resolution of 50 ps and a dead time of 77 ns. It generates an electrical pulse upon photon detection, which is sent to a time-correlated single-photon counting (TCSPC) module (PicoQuant PicoHarp 300). This module records the detection times of all photons and sync pulses, subsequently binning the photon arrival times based on their occurrence relative to a sync pulse. This process effectively employs a time-correlated detection technique akin to a lock-in amplifier, allowing us to identify optical changes synchronized with the sync pulse.

For the SOR measurements, the chopping signal is the sync pulse. For the FOR measurements, a 10 MHz signal phase-locked to the SAW frequency is used. All optical measurements are conducted at a SPAD count rate of 500 kCounts/s to ensure that the detected signal is not significantly distorted by photon pileup effects (56, 57).

The relative brightness at the NPoM location of each of the single wavelengths output by the supercontinuum laser is unknown. Thus, to normalize the time traces in Fig. S7 and Fig. S8 and create the measured 'true' time resolved spectrum of the NPoM of interest, we first find the 'true' spectrum averaged over many SAW periods $S_{avg,true}(\lambda)$ with a calibrated wideband light source, and use this to normalize the time traces. The 'true' spectrum of an NPoM when the SAW is on is found using the following equation:

$$S_{avg,true,on} = \frac{S_{NPoM,on} - S_{dark}}{S_{lamp} - S_{dark}} \quad ,$$

Eq. S4

where $S_{dark}$ is the dark spectrum of the spectrometer, $S_{NPoM,on}$ is the measured spectrum of the NPoM under lamp illumination when the SAW is on, and $S_{lamp}$ is the true spectrum of the lamp measured by placing a Lambertian scatter with wavelength independent attenuation in place of the sample.

Once the 'true' time-averaged spectrum is determined, we can determine the relative amplitude of each of the single-wavelength time traces in Fig. S7 and Fig. S8. For the SOR, we take the value of the time trace when the SAW has been on for a long period of time (right before it is switched off) and set that to the value of $S_{avg,true,on}$. For the FOR, we set the temporal average of the trace to be equal to the value of $S_{avg,true,on}$. By doing so, we adjust the amplitude of each time trace appropriately and stitch them together to produce the time-resolved spectrum shown in in Fig. 2.

Device fabrication.

The interdigitated Transducer (IDT) and adjacent metal patterns are constructed using an electron beam evaporation of a gold (Au) layer that is 150 nm in thickness. This layer is adhered to the substrate with either 5 nm of Chromium (Cr) or Titanium (Ti) and patterned onto a Y-cut Lithium Niobate (LiNbO$_3$) substrate, which is 500 μm thick. The patterning process employs direct write photolithography (Heidelberg MLA-150) and is completed using a single mask process. To create the patterns, two methods are used: metal lift-off and argon reactive ion etching (RIE). Some samples are produced using a lift-off process, while others are created with RIE. However, both methods yielded comparable results. Each entire device is then coated with a polymer stack of three different polymers: (i) an adhesion layer of (3-Mercaptopropyl)trimethoxysilane (MPTMS) to enhance adhesion between PDMS and Au; (ii) Sylgard 184 PDMS; and (iii) an adhesion layer of (3-Aminopropyl)triethoxysilane (APTES) to enhance adhesion to the AuNP. The MPTMS is deposited by soaking the device in 0.45% v/v MPTMS (purity 95%) in anhydrous ethanol for 60min (*58*). The PDMS layer is then made as in reference (*59*) by mixing Sylgard 184 at a ratio of 1:10 w/w and then diluting it to be 0.1-0.5% v/v in anhydrous toluene. The PDMS solution is spun onto the MPTMS coated samples at 6000 RPM for 60 s. Before the PDMS has time to cure, the APTES adhesion layer is deposited by soaking for 60 min in 1% v/v APTES (purity 99%) in deionized water (*60*). The triple stack is then baked at 100 °C for 120 min to cure the PDMS. Finally, colloidal 100-nm-diameter AuNPs capped with adsorbed ligands with carboxylic-acid functionalization (Nanopartz AC11-100-NPC-DIH) are drop-cast and left to sit for $10 - 200$ s, after which the sample is blown dry with a nitrogen gun. This process is outlined in Fig. S1.

Optical electromagnetic simulations.

The three-dimensional plasmonic structure is simulated using the finite element method (COMSOL radio frequency module). The model for the optical properties of Au is taken from Johnson and Christy (*61*) and the polymer stack is modeled as a uniform dielectric slab with a refractive index of 1.45 and thickness equal to the size of the gap (*62–64*). The 100 nm-diameter AuNP is modeled with a circular bottom facet. The scattering cross section $\sigma_{sc}$ is obtained by integrating the scattered fields from an incident plane wave on a surface that encloses the entire NPoM structure:

$$\sigma_{sc} = \frac{1}{I_0} \oiint\limits_{\Omega} (\boldsymbol{n} \cdot \boldsymbol{S_{sc}}) \, d\Omega \,, \qquad \text{Eq. S5}$$

where $I_0$ is the intensity of the incident light, $\Omega$ is some closed surface around the NPoM, $\boldsymbol{n}$ is the normal vector to this surface, and $\boldsymbol{S_{sc}}$ is the Poynting vector of the scattered fields.

Lumped mechanical model and vibrational mechanics of the NPoM

The mass-spring-damper Voigt model of the system, as shown in Fig. 4A of the main text, has an equation of motion:

$$m\ddot{g} + m\ddot{b} - F_{vis}(\dot{g}) - F_{spr}(g) = 0. \qquad \text{Eq. S6}$$

Here, $b = b(t)$ is the vertical displacement of the substrate, generating a driving force of $m\ddot{b}$. The viscous force $F_{vis}$ is a function of the rate at which the gap is changing $\dot{g}$, and the spring force $F_{spr}$ depends on the current gap size $g$ and the gap size when the SAW is off $g_0$. The gap $g = g(t)$ is the true size of the gap and is a function of time. The dot denotes a derivative with respect to time.

For a linear spring $F_{spr} = -k(g - g_0)$ and a linear asymmetrical damper $F_{vis} = -\zeta_{\pm}\dot{g}$ where $\zeta_{\pm}$ is given in Eq 1. Thus, the full equation of motion for a linear spring with an asymmetrical Newtonian damper is:

$$m\ddot{g} + m\ddot{b} + \zeta_{\pm}\dot{g} + k(g - g_0) = 0. \qquad \text{Eq. S7}$$

Using a vibrational mechanics analysis (*46*), we break this up into its fast and slow components by expressing $g$ as $g = G + \tilde{g}$, where $G$ is the slowly varying gap size, and $\tilde{g}$ is the rapidly oscillating component. We can now break up the equation of motion into two equations, one for the slow and one for the fast motion of the gap:

$$m\ddot{G} - \langle F_{vis}(\dot{g}) \rangle - \langle F_{spr}(g) \rangle = 0 \;, \qquad \text{Eq. S8}$$
$$m\ddot{\tilde{g}} + m\ddot{b} - F_{vis}(\dot{g}) + \langle F_{vis}(\dot{g}) \rangle - F_{spr}(g) + \langle F_{spr}(g) \rangle = 0 \;. \qquad \text{Eq. S9}$$

Where the bracket notation $\langle F \rangle$ denotes the time average over the temporal period of $\tilde{g}$. We see that if we add both equations, we recover the original equation of motion. Assuming that $\tilde{g}$ can be represented accurately by its first harmonic $\tilde{g} = \tilde{g}_0 \sin(\omega_{saw} t)$, we can define the time average as:

$$\langle F \rangle = \frac{1}{2\pi} \int\limits_0^{2\pi} F \, d(\omega_{saw} t) \qquad \text{Eq. S10}$$

Further, if we assume that the forces generated from the SAW and inertia of the AuNP are significantly larger than the other forces in Eq. S9 ($m\ddot{b} \gg F_{vis}$ and $m\ddot{b} \gg F_{spr}$), the equation for the fast motions simplifies to $m\ddot{\tilde{g}} = m\ddot{b}$. Thus, we find that $\tilde{g} = b_0 \sin(\omega_{saw} t)$. Since $\tilde{g}$ is now known, we can express the equation for the slow motions Eq. S8 as a function of $G$ alone:

$$m\ddot{G} - \langle F_{vis}(\dot{G} + \dot{\tilde{g}}) \rangle - \langle F_{spr}(G + \tilde{g}) \rangle = 0 \quad , \qquad \text{Eq. S11}$$

$$m\ddot{G} - \langle F_{vis}(\dot{G} + \omega b_0 \cos(\omega_{saw} t)) \rangle - \langle F_{spr}(G + b_0 \sin(\omega_{saw} t)) \rangle = 0 \quad . \qquad \text{Eq. S12}$$

Evaluating the time average of the viscous and spring forces using Eq. S10, we obtain a single differential equation that captures the dynamics of the slow gap $G$:

$$m\ddot{G} + \frac{1}{\pi}\left[ \dot{G}\left( \Delta\zeta \cos^{-1}\left( \frac{\dot{G}}{\omega b_0} \right) + \zeta_+ \pi \right) - \Delta\zeta b_0 \omega \sqrt{1 - \left( \frac{\dot{G}}{\omega b_0} \right)^2} \right] + k(G - g_0) = 0 \quad . \qquad \text{Eq. S13}$$

Where $\Delta\zeta = \zeta_- - \zeta_+$. This is a nonlinear differential equation, but since $G$ is the slow gap and varies slowly over the SAW period and $\dot{G} \ll \omega b_0$, we can linearize and simplify this equation to:

$$m\ddot{G} + \bar{\zeta}\dot{G} + kG = kg_0 + \frac{\Delta\zeta b_0 \omega}{\pi} . \qquad \text{Eq. S14}$$

Here, $\bar{\zeta} = \frac{\zeta_- + \zeta_+}{2}$. When we start at rest with $G(t = 0) = g_0$ and turn on the saw at $t = 0$, the general solution to this linear ordinary differential equation is:

$$G(t) = \frac{\Delta\zeta b_0 \omega \tau_1 \tau_2}{m\pi(\tau_1 - \tau_2)}\left\{ \tau_2 e^{-t/\tau_2} - \tau_1 e^{-t/\tau_1} \right\} + \frac{\Delta\zeta b_0 \omega}{k\pi} + g_0 . \qquad \text{Eq. S15}$$

Here, $\tau_1 = \frac{\bar{\zeta} + \sqrt{\bar{\zeta}^2 - 4mk}}{2k}, \tau_2 = \frac{\bar{\zeta} - \sqrt{\bar{\zeta}^2 - 4mk}}{2k}$. The numerical solution to Eq. S7 and the analytical approximate solution of Eq. S15 are plotted in Fig. 4D of the main text.

<u>Measurement of the device temperature during SAW operation</u>

To get an understanding of the expected temperature of the NPoM, we measure the temperature of a device using Raman thermometry while it is being excited with SAWs at different RF powers (Fig. S11). Raman thermometry is a temperature measurement technique that utilizes the relative intensities of the Stokes ($I_s$) and anti-Stokes ($I_{as}$) Raman peaks to determine the local temperature. The relative amplitudes of these peaks provide a measurement of the phonon population distribution at the location of interest, which is directly related to temperature via the Bose-Einstein distribution.

Phonons in the material interact with incoming photons, changing the energy of the photons by an amount equal to the energy of the phonons. The frequency shift resulting from this inelastic scattering indicates the phonon population at a specific energy. The scattered photons with altered energy are then detected by the Raman system as Stokes and anti-Stokes Raman peaks. The Stokes and anti-Stokes peaks may exhibit different optical coupling efficiencies, including effects such as Purcell enhancement, at the location of interest. Therefore, a constant correction factor $F_c$ must be calculated at a known temperature to account for these differences. The relationship between the temperature and the Raman peaks is given by (65, 66):

$$F_c \frac{I_{as}}{I_s} = \left(\frac{v_0 + v_R}{v_0 - v_R}\right)^4 \exp\left(-\frac{hv_R}{k_B T}\right). \qquad \text{Eq. S16}$$

Here, $v_0$ is the frequency of excitation of light, $v_R$ is the Raman band position (frequency of vibrational mode), $h$ is planks constant, $k_B$ is Boltzmann's constant, and $T$ is temperature.

Raman thermometry is conducted with a 0.9 mW 633 nm laser excitation of a section of LiNbO$_3$ between the IDT and metal film with NPoMs atop (Fig. S11A). The first three prominent Raman peaks at shifts of 150 cm$^{-1}$, 236 cm$^{-1}$, and 320 cm$^{-1}$ are used to find the temperature, and a measurement was done at room temperature to determine the correction factor $F_c$ for each of the peak pairs.

Although we measure a point on the LiNbO$_3$ substrate within 10 μm from the gold film and not directly the temperature of the polymer in the gap, thermal simulations show that this temperature is nearly the same as the temperature on the gold film itself. The results of these experiments are summarized in Fig. S11B. When 2 W of power is used to excite the device (which is the maximum power used in all experiments) the temperature rise above ambient is less than 60 °C.

Measurement of the polymer spacer thickness in the NPoM

The thickness of the polymer stack (consisting of MPTMS, PDMS, and APTES) is ascertained using ellipsometry (Woollam). Our model employs a two-layer system, consisting of an infinitely thick Au substrate with a layer of PDMS on top. Woollam-provided models for PDMS and Au are used. Only the thickness of the PDMS is fitted, while the index of refraction for the PDMS is not fitted.

For each measurement, a control device is fabricated. This device undergoes the same cleaning process as the coated devices but does not receive any polymer coating. The thickness of the measured "PDMS" layer on the control sample, usually between one and two nanometers, is subtracted from the thickness of the coated samples to estimate the total polymer coating thickness. The error margin for each measurement is set to be ± 1 nm. Here we assume that the loosely bound adsorbed ligands on the AuNP are displaced by the amine groups of the APTES and do not contribute to the total gap thickness (30, 67). Also, we assume the index of APTES and MPTMS is close enough in refractive index to PDMS so that the PDMS material model will accurately measure their thickness correctly (62–64). This is corroborated by the fact that when we use only a PDMS model to measure the thickness of each of these layers, we find that their thicknesses are approximately 1 nm each, which is expected for a monolayer of each.

Measurement of the slow optical response using a camera.

To quantify the magnitude of the slow optical response for multiple NPoM structures over a range of input RF frequencies or powers—as shown in Fig. 4E, Movies S1–S3, and Fig. S19—we use a charge-coupled device (CCD) camera array to efficiently collect extensive datasets. Videos are recorded using an unpolarized dark-field microscope while sweeping the frequency or power of the SAW. The SAW power is modulated at 1 Hz, and a lock-in technique is applied in post-processing to isolate the optical modulation signal at this frequency. The amplitude of this extracted signal represents the amplitude of the SOR. This approach enables simultaneous measurement of multiple NPoMs over a broad range of input powers or frequencies.

Use of ChatGPT for preparing the manuscript.

The paper does not contain any text generated by ChatGPT that was not originally drafted by the authors. However, ChatGPT models GPT-4, GPT-4o, o1, o2, and o3 were used to check for grammatical errors and partially refine the language. The writing workflow with ChatGPT was as follows: After a section was drafted by the authors, the text was fed into ChatGPT with specific instructions on how to identify and report errors. The output was then reviewed and edited by the authors to produce the final submitted text.

The physics and structure of all codes were developed by the authors themselves and ChatGPT was used for error correction, debugging, and pseudocode to source code translation. Buggy code along with the associated error messages were given to ChatGPT for debugging and the corrections were then reviewed and analyzed by the authors. Additionally, ChatGPT was used to translate pseudocode into source code: detailed plain-language instructions for specific code tasks were provided to ChatGPT, and ChatGPT transformed them into MATLAB source code according to those specifications. All final codes were reviewed and edited by the authors.

**Supplementary Text**

<u>S1. Thermal considerations.</u>

In the main text, we present one plausible model for the changes in color of the NPoMs during SAW excitation. However, we are interested in determining whether thermal effects could also play a role in generating the observed optical effects. In principle, there are many ways that large changes in temperature can produce the observed optical effects. These include: thermal expansion of the material in the gap, thermally induced stiffness changes of the material in the gap, changes in the refractive index of the Au and organic materials, and thermally induced diffusion and transport of materials. For this reason, it is important to identify the possible sources that can generate heat:

1. Mechanical damping of the SAW in the gold film
2. Mechanical damping in the polymer
3. Joule heating from the SAW-induced currents in the gold film
4. Optical heating resulting from the optical illumination
5. Thermoelastic effects

PDMS is remarkably stable over a wide range of temperatures around room temperature, and experiences very little change in its mechanical and optical properties (*68, 69*). Thus, a very significant rise in temperature would be needed to observe any significant optical modulation. Reportedly, the PDMS's visible refractive index varies by less than 2% over this temperature range (*68*). As for thermal expansion, PDMS has a coefficient of thermal expansion of 270 ppm/°C (*69*), indicating that it will expand less than 2% when raised to 60°C, which is significantly less than the observed strain of nearly 100%. Although the measured estimated temperature increase of 60°C is unlikely to cause significant optical modulation, we conduct a set of experiments and simulations to further argue that heat is not the cause of the slow and fast optical modulations.

To experimentally confirm that heat alone will not cause the optical shifts we observe, we place one of our devices on a hot plate, and raise the temperature to 120 °C. During this process, we monitor the spectral shifts of three NPoMs previously responsive to SAW activation and no significant spectral changes are observed in response to the applied heat (Fig. S15). Further, as detailed in sections S2 – S6, we estimate the spatial and temporal distributions of temperature in our device and correlate them with the spatial and temporal distributions of the observed optical effects. We find that the distributions of heat are not consistent with the distributions of the optical effects. Thus, we can reasonably conclude that heat is not the cause of the observed optical effects.

<u>S2. Spatial distribution of temperature at the edge of the collimated SAW beam.</u>

The SAW can generate heat via any of the aforementioned methods of heat generation. To determine how this heat spatially distributes throughout the chip, we first need to determine where the SAW is present, and then model how heat generated in this area diffuses to other areas of the device.

The SAW beam is well collimated as it propagates across the surface due to the IDT's wide aperture that is more than 50 times the SAW wavelength (*42*) Further, we can assume that the SAW propagates at least 1.25 mm into the metal plane area. This value is backed by SAW amplitude measurements, and the locations of NPoM that are experiencing scattering modulation. Thus, the SAW generated heat source has a well-defined edge, and extends at least 1 mm into the Au film region.

We can compute the temperature distribution of this SAW generated heat source by creating a finite element analysis (FEA) model of our device to numerically simulate how the temperature is distributed across the device surface in the steady state (Fig. S12). We create a full 3D model of a single device, including the substrate, thermal contact on the reverse side of the substrate with the holder, and the fabricated gold pattern on the top side (Fig. S12A,B). Heat is dissipated by external natural convection on the top surface, and through a piece of Kapton tape on a 1 oz copper PCB on the bottom surface. The copper clad PCB is modeled as a constant temperature boundary. The SAW heat generation is modeled as a 1W uniform heat source across the aperture of the IDT and extending about 1.25 mm into the Au film region.

We can see in Fig. S12C and Movie S2 that the NPoMs within the SAW beam area appear a different color and are experiencing the slow optical response, and NPoMs just off the SAW beam, within 10 μm from its edge, do not experience any optical modulation. However, when we map the temperature distribution at the edge of the SAW beam, we find there is only a ~5 °C difference in temperature across the edge of the SAW beam. Thus, the NPoMs just off the edge of the SAW beam and the NPoMs right inside the edge of the SAW beam experience roughly the same temperature. Therefore, the slow optical response cannot be caused by temperature effects.

<u>S3. Spatial distribution of heat in a standing SAW.</u>

In some of our experiments, we have notice apparent standing acoustic waves generated in our device, creating a noticeable interference pattern over the surface visible in the slow optical response of the nanoparticles (Fig. S21C, and Movie S3). We believe these interference patterns are generated by interference of the SAW propagating along the surface and a BAW that is reflected off the back of the chip. This is verified by the fact that when the back of the LiNbO$_3$ chip is roughened with sandpaper with an approximate 150 μm grain size (120 grit sandpaper) to create a SAW scattering interface on the back of the chip, the interference pattern disappears (Fig. S21).

Irrespective of the cause of the slow optical response periodic pattern, it shows the slow optical response can vary over the length of the wavelength of the SAW, which is approximately 5 μm at 700 MHz. In the following, we show both analytically and numerically that the temperature is nearly constant over this spatial period, and thus cannot be the cause of the observed optical modulation.

The heat equation in three dimensions in a uniform isotropic solid is:

$$c_p \rho \frac{\partial T}{\partial t} = \kappa \nabla^2 T + q_v \; .$$

Eq. S17

Here, $\rho$ is the mass density, $c_p$ is the specific heat capacity, and $\kappa$ is the thermal conductivity, $T$ is the temperature, and $q_v$ is a volumetric heat source. The SOR and the standing wave pattern is steady in time, and thus we solve the heat equation for the thermal steady state $\frac{\partial T}{\partial t} = 0$.

Consider a standing SAW that is propagating in the $\hat{z}$ direction, where $\hat{y}$ is normal to the surface and $\hat{x}$ is perpendicular to the direction of propagation. This standing SAW generates heat over a surface based on the amplitude of the standing wave across the surface. We assume a wave with amplitude $\sim \cos(kz)$ will generate heat at a rate that is proportional to the square of the amplitude $\cos^2(kz)$ (70). Thus, the heat generated from a standing SAW averaged over the SAW temporal period is $q_{saw} = 2\langle q_{saw} \rangle \cos^2(kz)$ where $\langle q_{saw} \rangle = \frac{2\pi}{k} \int_0^{2\pi/k} q_{saw} dz$ is the heat generation rate per unit volume averaged over space and over the SAW temporal period.

In the steady state, when $\frac{\partial T}{\partial t} = 0$, the total heat deposited into a system must be the same as the heat leaving the system. If we assume that the cooling rate $q_c$ is roughly constant over the surface and independent of temperature (which is true if the temperature variation across the surface is small, as we will show), in the thermal steady state the total heat into and out of the device must add to zero:

$$0 = \int_V (q_{saw} - q_c) dV$$

Eq. S18

$$0 = \langle q_{saw} \rangle V - q_c V$$

Eq. S19

$$q_c = \langle q_{saw} \rangle$$

Eq. S20

Now we express the spatially-dependent total heat generation rate in the metal as the heat generated by the SAW and the heat taken away by cooling:

$$\begin{aligned} q_v &= 2\langle q_{saw} \rangle \cos^2(kz) - q_c \\ &= 2\langle q_{saw} \rangle \cos^2(kz) - \langle q_{saw} \rangle \\ &= \langle q_{saw} \rangle (2\cos^2(kz) - 1) \\ &= \langle q_{saw} \rangle \cos(2kz) \end{aligned}$$

Eq. S21

Where $q_v$ is the heat source per unit volume in the thermal steady state. Further, we assume that the thermal conductivity of LiNbO$_3$ is negligible compared to that of gold ($\kappa_{LN} = 5 \frac{W}{m \cdot K}$ and $\kappa_{Au} = 315 \frac{W}{m \cdot K}$), and thus the heat transfer is performed solely by the gold film and the top and bottom of the gold are thermal insulators. In addition, the SAW wavelength is much larger than the Au film thickness, and the electrical skin depth of Au at the SAW frequency is much larger than the film thickness. For these reasons, we can assume that the heat is uniformly generated over the thickness of the film and $\frac{\partial^2 T}{\partial y^2} = 0$. Additionally, since the acoustic beam width is much larger than the SAW wavelength, we can assume that $\frac{\partial^2 T}{\partial x^2} = 0$. Thus, the heat equation for our system becomes:

$$\frac{\partial^2 T}{\partial z^2} = \frac{1}{\kappa_{Au}} q_v \ .$$

Eq. S22

By substituting our value for $q_v$ expressed in Eq. S21 and solving for a temperature distribution of the form $T(z) = \Delta T \cos(2kz) + T_0$ where $T_0$ is some constant that represents the average temperature across the surface, we find:

$$\Delta T = \frac{\langle q_{saw} \rangle}{4\kappa_{Au} k^2}.$$

Eq. S23

Where $T_0$ is some constant that represents the average temperature across the surface, and $\Delta T$ is the amplitude of the spatially-varying temperature that we are looking for. For a $k = \frac{2\pi}{10[um]}$ and $\langle q_{saw} \rangle = 5 \cdot 10^{12} \ W/m^3$ (which is the heat source generated by 1 W of power being absorbed into a 3 mm x 0.5 mm x 150 nm piece of material) we find that the spatial variation of temperature over the standing wave pattern is:

$$\Delta T \approx 0.01 \ K.$$

This value is supported by a FEM heat transfer simulation shown in Fig. S13. We expect the mechanical and optical properties of our device to remain constant over such a small temperature variation. Thus, the SOR wave pattern we notice in Fig. S21 and Movie S3 cannot be attributed to temperature variations across the surface.

### S4. Temporal dependence of the thermal response and its possible connection to the slow optical response.

We observe in our measurements that the slow response has a rise time on the order of 10 μs, and a similar fall time (Fig. 3). Here, we compare that result to the simulated thermal rise time. In the same simulation geometry and thermal conditions as described in Fig. S12A and section S2, we now perform a time-dependent thermal simulation where the heat source initially is turned off, and then is immediately switched on while the temperature on the metal plane adjacent to the IDT is measured. The results are summarized in Fig. S12E. As we see from the figure, the rise and fall time of the temperature on the surface in response to SAW heating is on the order of 10 ms, which is more than three orders of magnitude longer than the SOR. Thus, it is reasonable to conclude that the SOR cannot be attributed to SAW-induced temperature changes.

### S5. Temporal dependence of the thermal response and its possible connection to the fast optical response.

As the SAW displacement field oscillates at angular frequency $\omega_{SAW}$, at a given point on the surface, it generates heat by both Joule heating and mechanical losses in the gold film. To calculate the time-dependent heat generation from Joule heating, we start with Joule's first law $q_v = \boldsymbol{J} \cdot \boldsymbol{E}$. In combination with Ohm's law we get:

$$q_v = \sigma_{Au} |\boldsymbol{E}(t)|^2 \ .$$

Eq. S24

Here, $q_v$ is the volumetric heat source in $W/m^3$, and $\sigma_{Au}$ is the conductivity of gold. The electric fields are generated by the SAW or RF source and oscillate at the SAW frequency:

$$E(t) \sim \sin(\omega_{SAW} t) \ .$$

Eq. S25

Thus, the heat generated from Joule heating is:

$$q_v \sim \sin^2(\omega_{SAW} t),$$
$$\sim 1 - \cos(2\omega_{SAW} t).$$

Eq. S26

Thus, Joule heating occurs at twice the SAW frequency. Similarly, mechanical damping also causes heating at twice the SAW frequency. However, from Fig. 3C, we can see that the FOR occurs at a rate $\omega_{SAW}$. Thus, it is not possible for the Joule heating or mechanical damping to drive the fast optical response.

## S6. Heating from mechanical damping in the gap of an NPoM.

Besides a solely thermal effect, there can be a thermal-mechanical cause for the optical modulations, where mechanical damping caused by small strain in the gap material causes heating only in the gap region. In this special case, fast mechanical deformations that are associated with the fast mechanical response slowly heat the gap material, expanding it and thus causing the slow optical response. Here we calculate the upper limit for the temperature rise by this mechanism. We first theoretically calculate the expected maximum amount of power dissipation in the gap from mechanical damping, and then use a heat-transfer COMSOL model to estimate the resulting gap material's temperature change produced by this thermal source. Without assuming anything about the material characteristics of the polymer, we can estimate the amount of heat generated. First, since we observe a frequency-independent, broadband mechanical response, we can assume that we are operating above the mechanical resonance frequency, where the gap size change is broadband and the spring forces are relatively small compared to the inertial forces. The viscous force on the dashpot is $F_{vis} = -\zeta \dot{g}$ and the power dissipated by it is $P = \zeta \dot{g}^2$. Since we are assuming small spring forces, the only force on the dashpot is the inertial force of the AuNP, which has a maximum of $F_{iner,max} = m|\ddot{x}|$. Therefore, the maximum power generated for the system is:

$$P_{max} = \zeta |\dot{g}|^2 = |F_{vis}| \, |\dot{g}| \leq m|\ddot{x}||\dot{g}| \, . \qquad \text{Eq. S27}$$

The location of the NP is $x = b + g$, where $b(t) = b_0 \cos(\omega t + \phi)$ is the driving displacement, $g = G + \tilde{g}$ is the total gap size, $G$ is the slow moving gap size, $\tilde{g} = \tilde{g}_0 \cos(\omega t + \theta)$ is the quickly moving gap size, and $\theta$ and $\phi$ are arbitrary phases. Thus, immediately we see that $\ddot{x} = \ddot{b} + \ddot{G} + \ddot{\tilde{g}}$ and $\dot{g} = \dot{\tilde{g}} + \dot{G}$. For the fast mechanical response, we assume that both the speed and acceleration during the fast mechanical response are much greater than during the slow mechanical response, so $|\ddot{G}| \ll |\ddot{\tilde{g}}|$ and $|\dot{G}| \ll |\dot{\tilde{g}}|$. From our experiments (Fig. 3F), we find that $\tilde{g}_0 \approx b_0$. Plugging in these values we find that $|\ddot{x}| = 2\omega^2 b_0$ and $|\dot{g}| = \omega b_0$ and:

$$P_{max} \leq 2m\omega^3 b_0^2 = 160 \, nW \, . \qquad \text{Eq. S28}$$

Using COMSOL's module for heat transfer in solids, we compute the expected temperature rise in the gap from a pulsed heat source with instantaneous power generation $P(t) = 2P_{max} \cos^2(\omega t)$, where $P_{max} = 160 \, nW$ and $\omega = 2\pi \cdot 700 \, MHz$. The result is shown in Fig. S14. Although the decay time of the temperature is like that of the slow optical response, the temperature rise is less than 5 °C. Such a small temperature rise cannot be expected to cause any significant optical changes.

## S7. Electrostatic force on the AuNP.

Another potential cause for the optical modulations is the force generated by the electric field of the SAW. Due to the piezoelectric properties of LiNbO₃, the mechanical displacement of the substrate is coupled to an electric field (Fig. S16). After fabrication and left to sit in ambient for long periods of time, the AuNP are likely charge neutral, but there will still be a force generated by the interaction of the polarizable AuNP dipole moment $\boldsymbol{p}$ with the gradient of the electric field. This force will be $\boldsymbol{F} = (\boldsymbol{p} \cdot \nabla)\boldsymbol{E} = \frac{1}{2}\alpha \nabla \left( \left\| \boldsymbol{E} \right\|^2 \right)$ where where $\alpha$ is the polarizability of the AuNP (71). The electric fields are $E_z \cos(k_{saw}z - \omega_{saw}t) \exp(-ay)$ and $E_y \sin(k_{saw}z - \omega_{saw}t) \exp(-ay)$ where $a \approx 1 \, \mu m^{-1}$, and will pull the AuNP down and side-to-side.

If we assume the maximum force case where the electric field is not screened at all by the Au film, and the polarizability of the AuNP is not screened either so that $\alpha \approx 4\pi\epsilon_0 r_{NP}^3$ (72), the total force $\left\| \boldsymbol{F} \right\|$ is limited to about 50 fN, which is already more than three orders of magnitude less than the expected mechanical forces. However, the Au film under the AuNP acts as a screening plane and also changes the mode shape of the SAW significantly, reducing the electric field at the surface (42). Thus, we expect the actual electrostatic forces to be much lower than this.

In the MHz range, the Au film under which the SAW travels is mechanically lossy and considerably attenuates the SAW as it travels across the surface. Further, as the SAW arrives at the rough edge of the LiNbO₃ substrate, the edge itself provides a sharp acoustic discontinuity and scatters much of the impinging SAWs into bulk acoustic waves (BAWs). Both effects combined severely limit the amplitude of reflected SAWs and prevent standing SAWs from being formed on the metal plane.

To experimentally corroborate these assertions, we find that the NPoMs are not observed to change scattering near the end of the Au film, even when they are changing their scattering properties near the IDT, thus indicating the SAW is mostly attenuated by the time it reaches the end of the Au film. Furthermore, the standing wave patterns visible in Fig. S21 disappear after eliminating BAWs by sanding the back of the substrate. This suggests that standing wave patterns are not generated by SAWs alone, and thus SAWs are best represented as traveling waves. Finally, adding acoustic dampers, such as silicone tape before the edge of the substrate, does not change any of the observed effects.

## S9: Long-term stability and repeatability of results.

Here we discuss the longevity of the device, and how stable and repeatable our results are over time. Immediately following the fabrication process, we observe a significant number of NPoM structures undergoing a color change on the surface. For polymers with a thickness greater than ~6 nm, approximately 95% of NPoMs exhibit a change, while for those less than 5 nm thick, the proportion is often around 20%. Over the subsequent days, some NPoMs cease to respond to the SAW, reducing these figures slightly. We attribute this to the slow creep and flow of the polymer out of the gap driven by the attractive surface forces between the AuNP and the metal film, which draw the two components closer together. In Fig. S20, we show that over 93% of the NPoMs initially responsive to the SAW remain responsive after over 1 year of aging and 5 trillion acoustic cycles.

Over a period of two years, we successfully fabricated more than ten distinct devices demonstrating observable optical modulation of NPoMs. The manufacturing process proves to be highly reproducible, consistently yielding functional devices with a yield rate that is limited by the reliability of the photolithography process.

## S10: Linear mechanical properties of the NPoM and comparison to expected values.

Here we compare the mechanical dynamics and material properties that we estimate from our measurements and mechanical model to the expected results from literature. The dynamics of our system are governed by both the viscoelastic material properties as well as the geometry of the NPoM structure. The structure of the NPoM system is estimated—as discussed in the main text—by polymer thickness measurements with ellipsometry and matching facet sizes between experiment and simulation using the scattering spectra. In our experiments, we observe that there is a lower limit to the frequency at which optical modulation occurs. This helps us determine the approximate resonant frequency of the NPoM system, and thus the spring constant in the initial state $k_0$.

Our results and theory indicate that the system is linear in two regimes: the unexcited state and the excited state. In the unexcited state, the SAW may be present, but it does not change the size of the gap. The NPoM thus follows the motion of a spring-mass system below resonance, and the change in gap is nearly zero. In the excited state, there begins to be a relative motion between the NP and the Au film, causing motion of the material in the gap and driving it to the excited state where $g = \tilde{g} + \Delta G$. The transition from the unexcited to the excited state requires some motion in the gap material, and we therefore expect it to happen near the resonant frequency of the NPoM in the unexcited state. This is because below the resonant frequency of a spring-mass-damper system there is no significant motion of the spring. The resonant frequency in the initial state can be estimated by $f_{0,0} = \frac{1}{2\pi}\sqrt{k_0/m}$, where $k_0$ is the spring constant in the unexcited, initial state when the NP is closest to the substrate and the gap size is $g = g_0$. In our experiments, we first start to see NPoMs get excited and start changing color around 400 MHz (Fig 4E, Fig. S18, Fig. S19). Setting this as the resonant frequency $f_{0,0}$, and setting the mass of the NP to that of a 100 nm diameter AuNP, we find that $k_0 \approx 63 \ N/m$.

To estimate the expected spring constant in the unexcited state, we look to the mechanical properties of PDMS available from literature and input these properties into a solid mechanics COMSOL model to determine the resulting spring constants. This axially symmetric, continuum mechanics COMSOL model determines the initial spring constant by applying a small force to the NP in the vertical direction and recording its resultant displacement. The resulting spring constant is then calculated as the force applied divided by displacement. The gap polymer is not entirely PDMS, and it has APTES and MPTMS ligands included as well, but we model it as PDMS here. Although the mechanical properties of PDMS have not been directly quantitively determined at frequencies as high as in our experiments, the principle of time-temperature superposition has been used to estimate the Young's modulus of PDMS at frequencies from mHz to GHz, plotted in Fig. S17A (37). Separately, to approximate this geometric contribution to the effective spring constant of the polymer layer, we create a solid mechanics model with different facet and gap sizes. The facets are modeled as circular, and the gap is modeled as an infinite elastic sheet of thickness $g$ and Poisson ratio $v = 0.49$ that resides between the infinitely stiff AuNP and Au substrate. We plot the effective spring constant independent of elastic modulus for different gap geometries in Fig. S17B. Finally, we combine both the material properties from literature and the geometric factors from our model and estimate the expected spring and drag constants. At 500 MHz, the Young's modulus of PDMS is approximately 20.6 MPa, and our FEM predicts that the ratio of $k/E$ in the initial satate (when the gap size is 4 nm and facet size is 40 nm) is 2.23 μm. Thus, the expected value of $k_0 = 46\ N/m$ is the product of these two. This is in close agreement with our value of 63 N/m estimated from the response to SAW.

S11: Lumped mechanical model including a nonlinear spring.

Although having a linear spring and asymmetrical dashpot can replicate the SMR and FMR quite well at a single acoustic frequency and amplitude, when we sweep the acoustic power (as shown in Fig. 4E, Fig. S18, and Fig. S19), we clearly see evidence that the effective spring is nonlinear. Below a certain threshold SAW amplitude, SOR is not observed for any NPoM (Fig. 4E, blue curve). Furthermore, when examining the SOR of individual NPoMs, the SOR does not appear proportional to SAW amplitude (see Fig. S18D, Fig. S18F, and Fig. S19A). Rather, the SOR appears to be zero below some threshold SAW amplitude, and then when this threshold is reached, the SOR substantially increases. These results are not possible with a linear spring as in Eq. 3.

To better understand and capture these dynamics with a lumped model, we consider the impact of introducing nonlinearity into the spring in the model as well. The spring constant is now a function of the gap size so that $\frac{\partial F_{spr}}{\partial g} = -k(g)$. There are a variety of nonlinear springs $k(g)$ that we can choose to model our system, but ultimately the model should capture the behavior that the spring stiffens for smaller gaps and softens for larger gaps (45-48, Fig. S17). Furthermore, for very large stretches, the polymer chains become fully extended and the entropy that contributes to the elastomer's softness is significantly reduced. Thus, we would expect the stiffness of the nonlinear spring to increase again at large gaps. An example of such a nonlinear spring and the potential it creates is presented in Fig. S19D. For this spring, the spring constant initially decreases as the gap increases, then increases again at large gaps.

With a nonlinear spring, we see a nonlinear change in the $\Delta G$ in response to a linear increase in $b_0$. To show this, let us now transform the approximate equation of motion for the slow gap that we derived for a linear spring to the nonlinear case. Eq. S14 becomes (under the first order approximation)

$$m\ddot{G} + \bar{\zeta}\dot{G} - F_{spr}(G) = \frac{\Delta\zeta b_0 \omega}{\pi}\ .$$

Eq. S29

At equilibrium, $\dot{G} = \ddot{G} = 0$ and we have $-F_{spr}(G) = \frac{\Delta\zeta b_0 \omega}{\pi}$. Taking the derivative of the equilibrium spring force $\frac{dF_{spr}}{db_0} = \frac{dF_{spr}}{dG} \cdot \frac{dG}{db_0} = \frac{-\Delta\zeta\omega}{\pi}$ where $\frac{dF_{spr}}{dG} \equiv -k(G)$, we can find the change in the equilibrium gap as the SAW power changes:

$$\frac{dG}{db_0} = \frac{\Delta\zeta\omega}{\pi k(G)}$$

Eq. S30

Therefore, we can explain the nonlinear behavior as we sweep $b_0$ by introducing a nonlinear spring $k(G)$. The slope of the slow gap increase with increasing SAW amplitude is simply inversely proportional to the nonlinear spring constant.

<u>S12: Analysis to show that the deformation in gold in the NPoM is negligible.</u>

In our mechanical model, we represent the entire NPoM system using a single spring to represent the polymer stiffness, assuming that the stiffness of the Au structure is significantly higher and can be considered infinitely stiff in comparison. To validate this assumption, we calculate the stiffness of the polymer gap relative to the surrounding Au. As forces near the gap increase, the polymer gap will deform, but the Au in the nanoparticle and the Au substrate also experience deformation. Following the analysis in reference (*35*) and section S11, we model these stiffnesses with springs with constants $k_{polymer}$, $k_{NP}$ and $k_{sub}$ for the polymer spacer, Au nanoparticle, and Au substrate respectively. In our system, which has a facet diameter of ~40 nm, a gap size of ~4 nm, a PDMS Young's modulus of ~20 MPa (*37*), a modulus of Au of 80 GPa, we find that the spring constants for the three deformable interfaces are $k_{polymer} = 63\ N/m$, $k_{NP} = 8600\ N/m$, and $k_{sub} = 4000\ N/m$. The resonant frequencies of each of these springs supporting a 100 nm AuNP as a mass is $f_{0,polymer} = 400\ MHz$, $f_{0,NP} = 4.6\ GHz$, and $f_{0,sub} = 3.2\ GHz$. Therefore, at the frequencies of our experiments, we expect negligible deformations in the Au itself relative to that of the polymer.

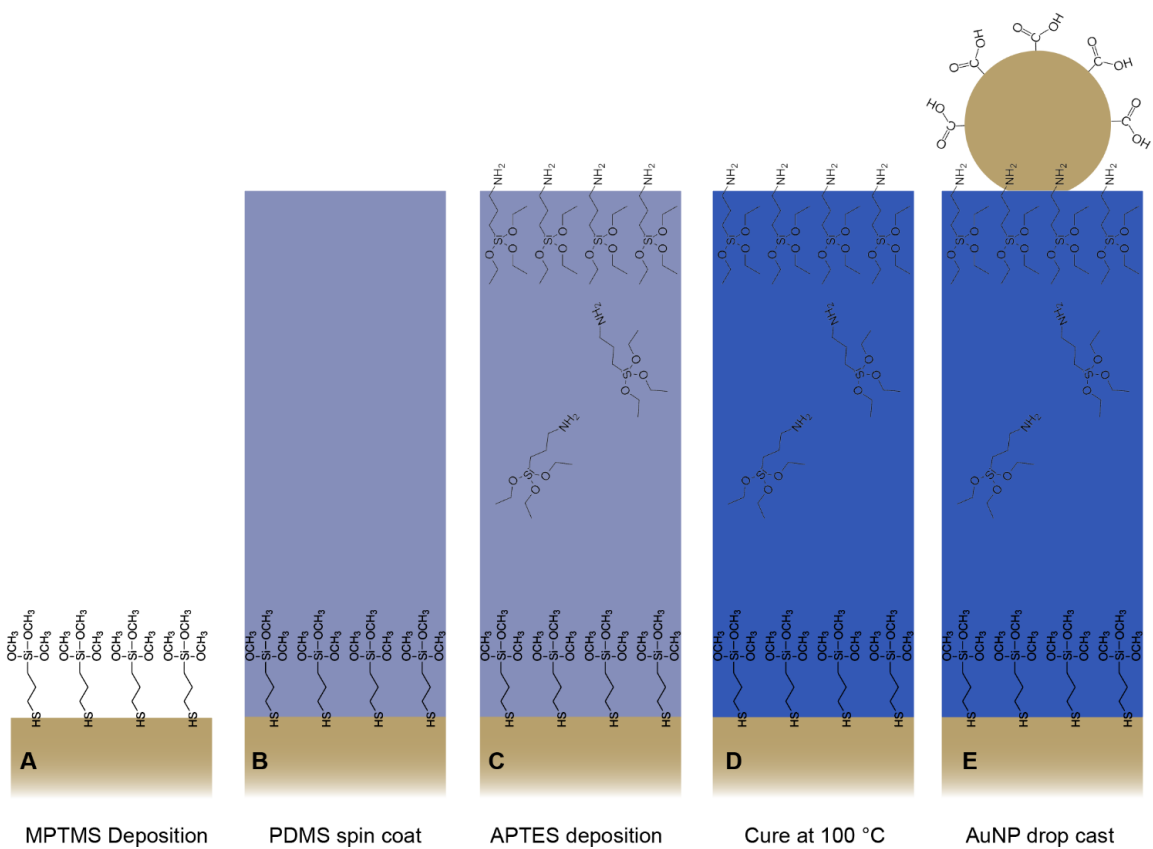

**Fig. S1. Nanoparticle-on-mirror fabrication process.**
(**A**) Deposit MPTMS by soaking the device in an MPTMS solution; this adhesion layer allows PDMS to adhere to the Au surface. (**B**) Spin-coat a highly diluted PDMS layer to act as a mechanically compliant plasmonic gap filler, and immediately (**C**) deposit APTES by soaking the device in solution for 1 hour; APTES facilitates adhesion between PDMS and the drop-casted AuNP. (**D**) Cure the PDMS in an air oven at 100 °C for 2 hours. (**E**) Finally, drop-cast carboxylic acid-functionalized 100 nm diameter AuNPs onto the surface.

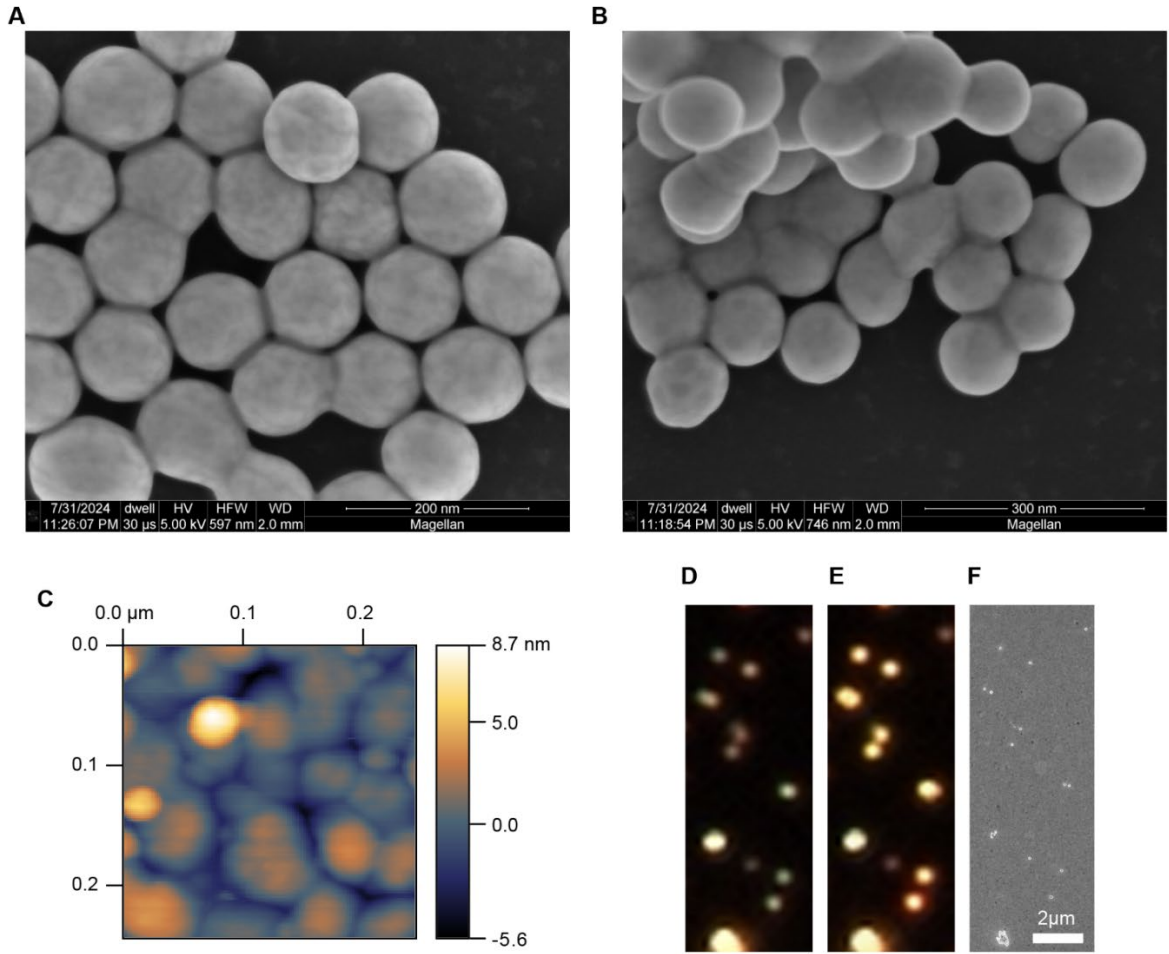

**Fig. S2. Scanning electron microscopy (SEM) and optical images of gold nanoparticles (AuNPs).**
(**A-B**) Scanning electron microscope images of 100 nm diameter AuNPs representative of those in all experiments. The sample is prepared by drop-casting AuNPs on a low-resistivity Si substrate, washing with water, and then plasma cleaning in the microscope chamber. The AuNPs are clearly faceted, and often have oblong shapes and flat sides. (**C**) Atomic force microscope images of a bare Au surface representative of those used in the experiments. The surface roughness has a root-mean-square value of 1.74 nm. (**D-E**) Dark-field optical images of the NPoMs shown in two states: (D) prior to the activation of the SAW, and (E) during the SAW activation. (**F**) An SEM micrograph of the same area as presented in (D) and (E), showing that both isolated NPoMs and those near others exhibit optical modulation.

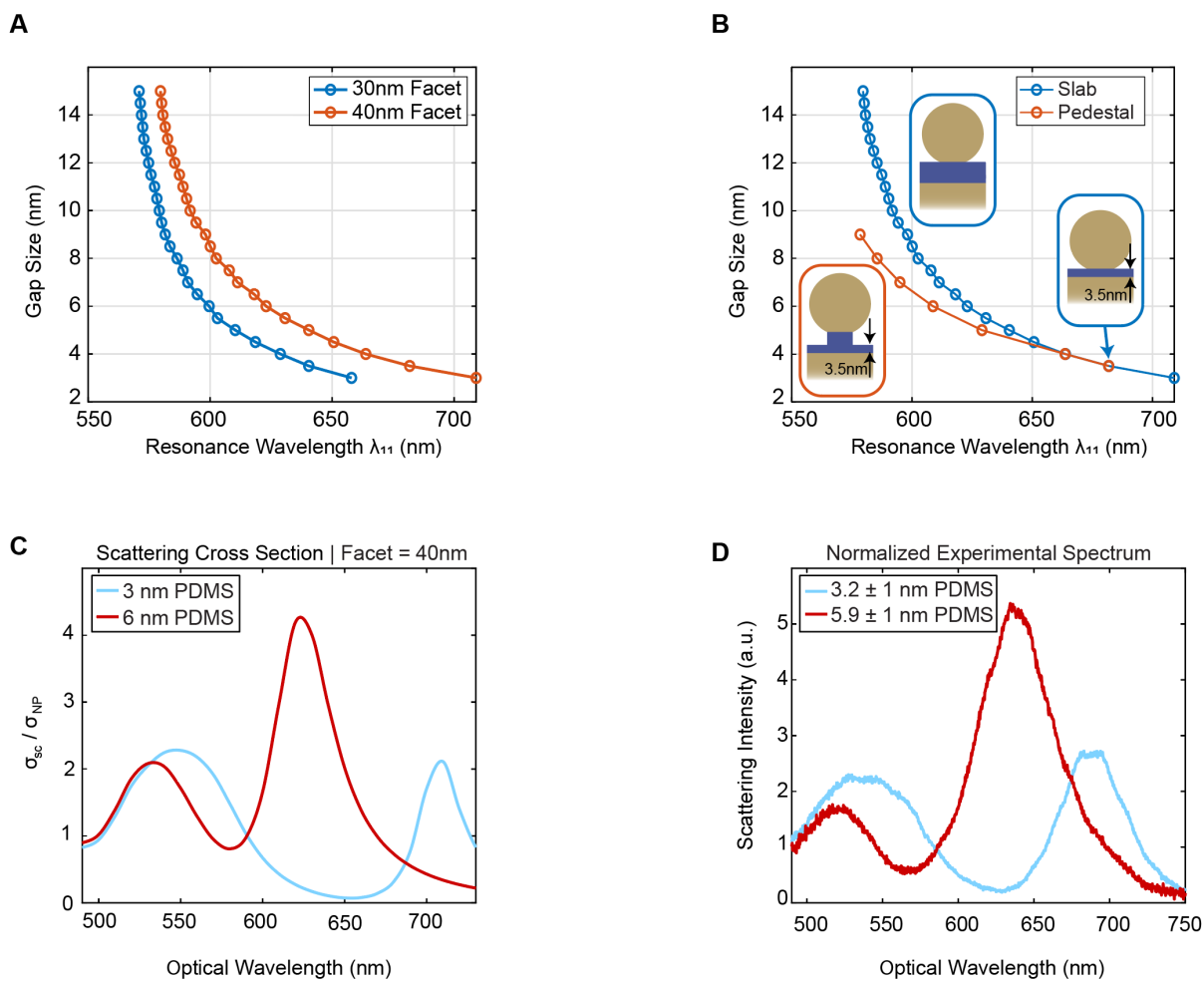

**Fig. S3. Study of NPoMs with different geometries.**
(**A**) Simulation of an NPoMs with a 30 and 40-nm-diameter circular facets. The fundamental mode red-shifts as the facet size is increased from 30 to 40 nm. (**B**) Simulations of NPoMs with 40 nm circular facets and different polymer conformations in the gap region. The blue curve shows the primary plasmon resonance for a uniform polymer slab. The orange curve shows the resonance for a polymer "pedestal," which has a 3.5-nm-thick foot and a circular shaft matching the facet diameter. We can see that the reduced amount of polymer in the gap blue shifts the spectrum with a pedestal. (**C**) NPoM scattering spectrum with facet size of 40 nm and polymer slab gap sizes of 3 and 6 nm. (**D**) Representative experimental scattering spectrum of two NPoMs on devices with measured polymer thicknesses approximately equal to that of the simulated thicknesses in (C). Normalization is performed by setting the spectral values at 500 nm to be equal.

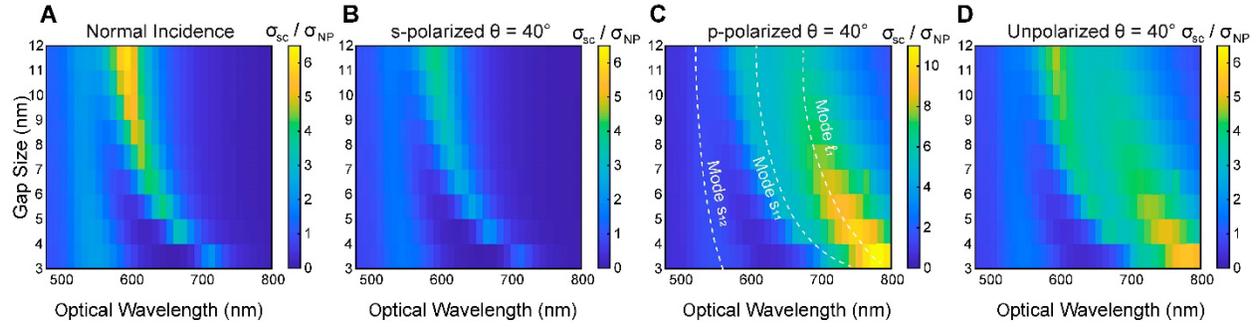

**Fig. S4. S-polarized and unpolarized scattering cross section of NPoM.**
The normalized scattering cross section $\sigma_{scat}/\sigma_{NP}$ is shown as a function of optical wavelength ($\lambda$) and gap size for (**A**) normal incidence, (**B**) s-polarized light at a 40° angle of incidence, (**C**) p-polarized light at a 40° angle of incidence, and (**D**) unpolarized light at a 40° angle of incidence. For s-polarized light, the shape of the scattering cross section remains unchanged with the angle of incidence, but the amplitude decreases roughly in proportion to $\cos\theta$, where $\theta$ is the angle of incidence. For p-polarized light, a vertical "antenna" mode $\ell_1$ is excited due to a component of the electric field perpendicular to the substrate, leading to a broadening of the scattering cross section. For unpolarized light, both s- and p-polarization effects contribute. The time-resolved spectra of this paper are taken with s-polarized light. However, some other experiments use unpolarized light, and the influence of the vertical mode can affect the measured spectra.

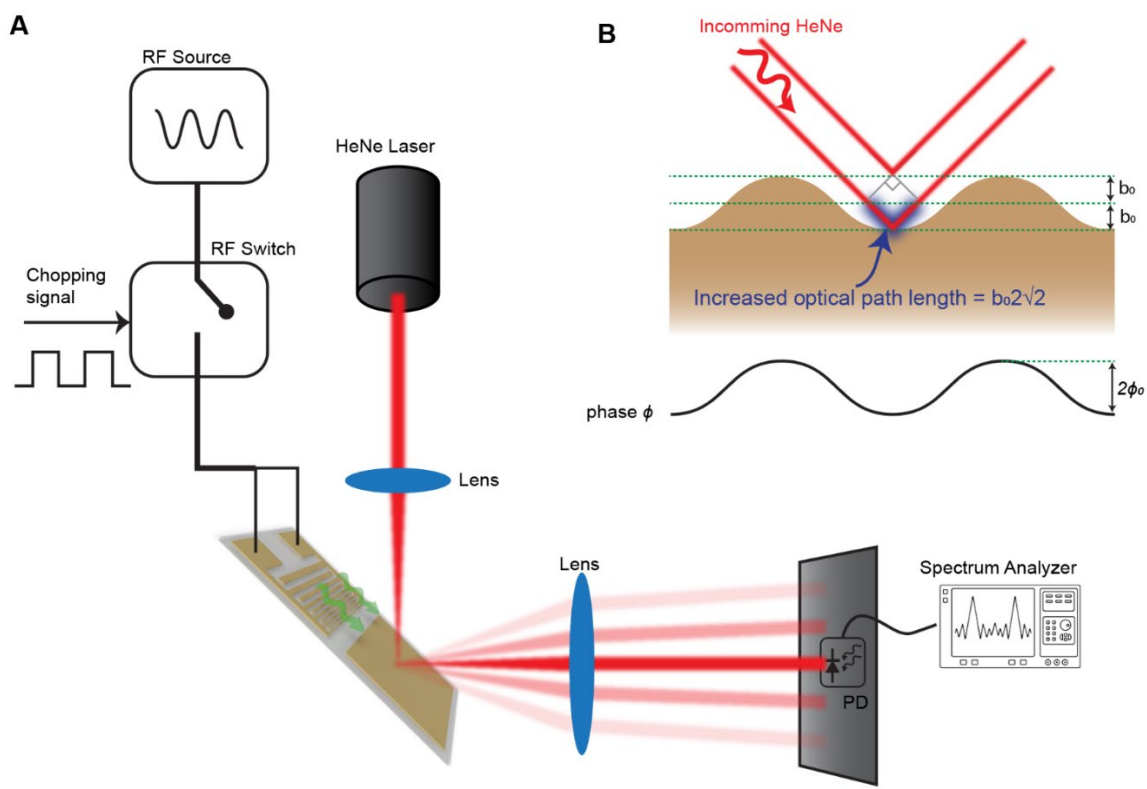

**Fig. S5. Measurement setup to quantify SAW-induced surface displacements of a thin metal film.**
(**A**) Setup for measuring the vertical displacement induced by a SAW in a thin Au film deposited on a LiNbO$_3$ substrate. A laser illuminates the surface of the Au film without any NPoM present while the SAW is amplitude modulated at 150 kHz to facilitate lock-in detection. The laser then reflects from the device and diffracts some of the incident power into higher-order modes due to the periodic surface undulation caused by the SAW. A photodiode is placed to measure the power in the 0$^{th}$ diffracted order to detect the amount of power lost to higher-order beams, and the signal from the photodiode is sent to a spectrum analyzer tuned to the modulation frequency. (**B**) Diagram of optical path length difference and resulting phase contrast caused by the SAW with a vertical displacement amplitude $b_0$.

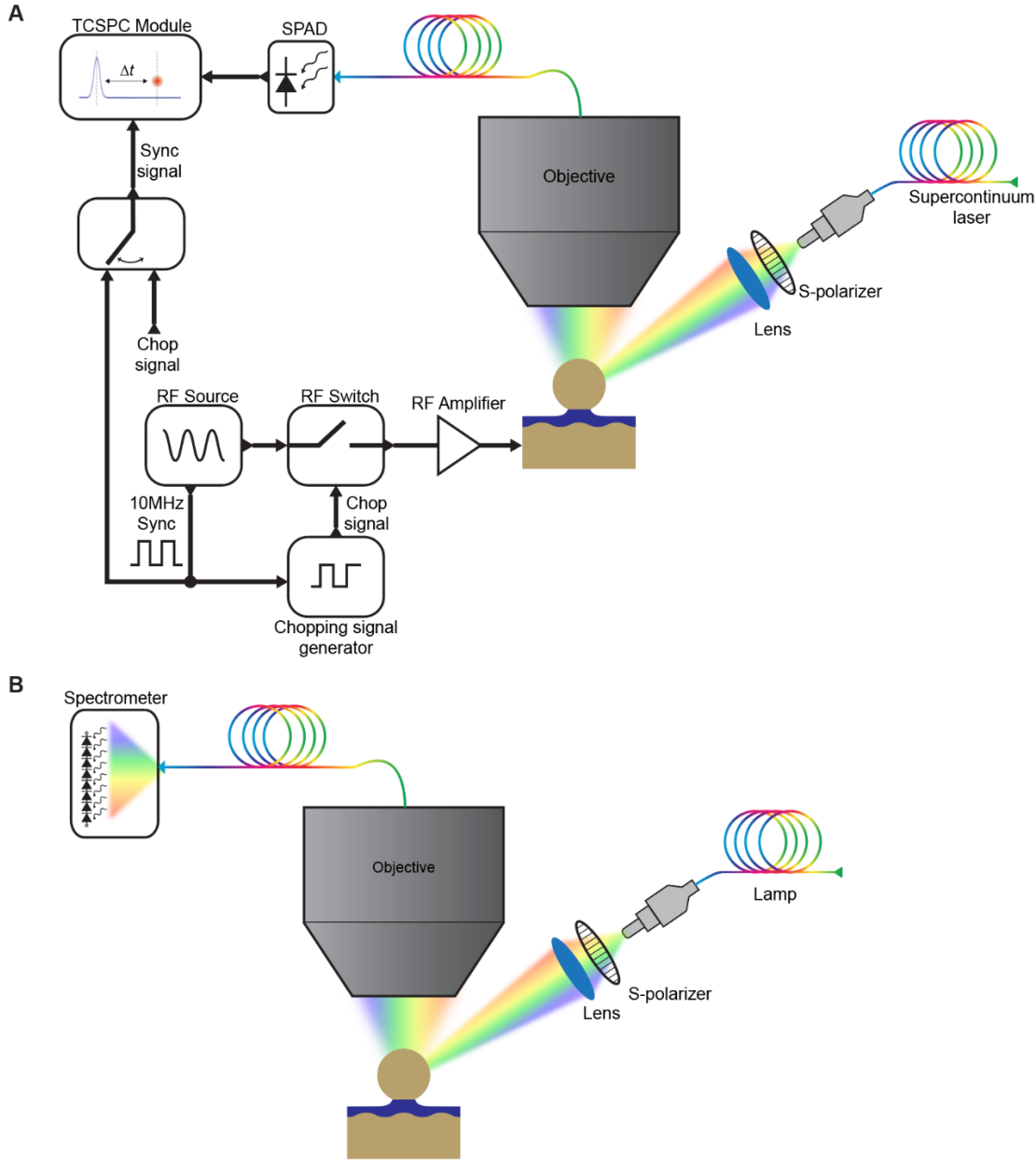

**Fig. S6. Setup to measure the time-resolved darkfield scattering spectra of individual nanoparticles.**
Optical paths are depicted in rainbow colors, and optical fibers are shown as curled lines. Electrical signals are shown as black lines. The time-resolved measurements are done as shown in (**A**): A single wavelength is filtered out from a supercontinuum laser, s-polarized, and focused onto the NPoM structure. The scattered light from the NPoM is collected by a microscope objective, and the microscope fiber-couples the light only from a single diffraction-limited spot where the NPoM structure resides. This outcoupled light is sent to a SPAD, the SPAD converts each photon into an electrical pulse, and the TCSPC module detects and timestamps these pulses with picosecond precision. Simultaneously, the TCPSC module timestamps the sync events. For measuring the FOR, the sync events are a 10 MHz signal that is phase-locked to the SAW driving signal. For measuring the SOR, the sync signal is the "chopping" signal that periodically turns on and off the SAW. Then, a new wavelength is filtered out from the supercontinuum laser and the measurement is repeated for wavelengths across the visible spectrum. Finally, the relative weights of those time traces are found by taking the time-averaged spectrum of the NPoM using the setup shown in (**B**), where the filtered laser is replaced with a broadband lamp, and the SPAD is replaced with a spectrometer. The time traces are then stitched together using these weights in post-processing.

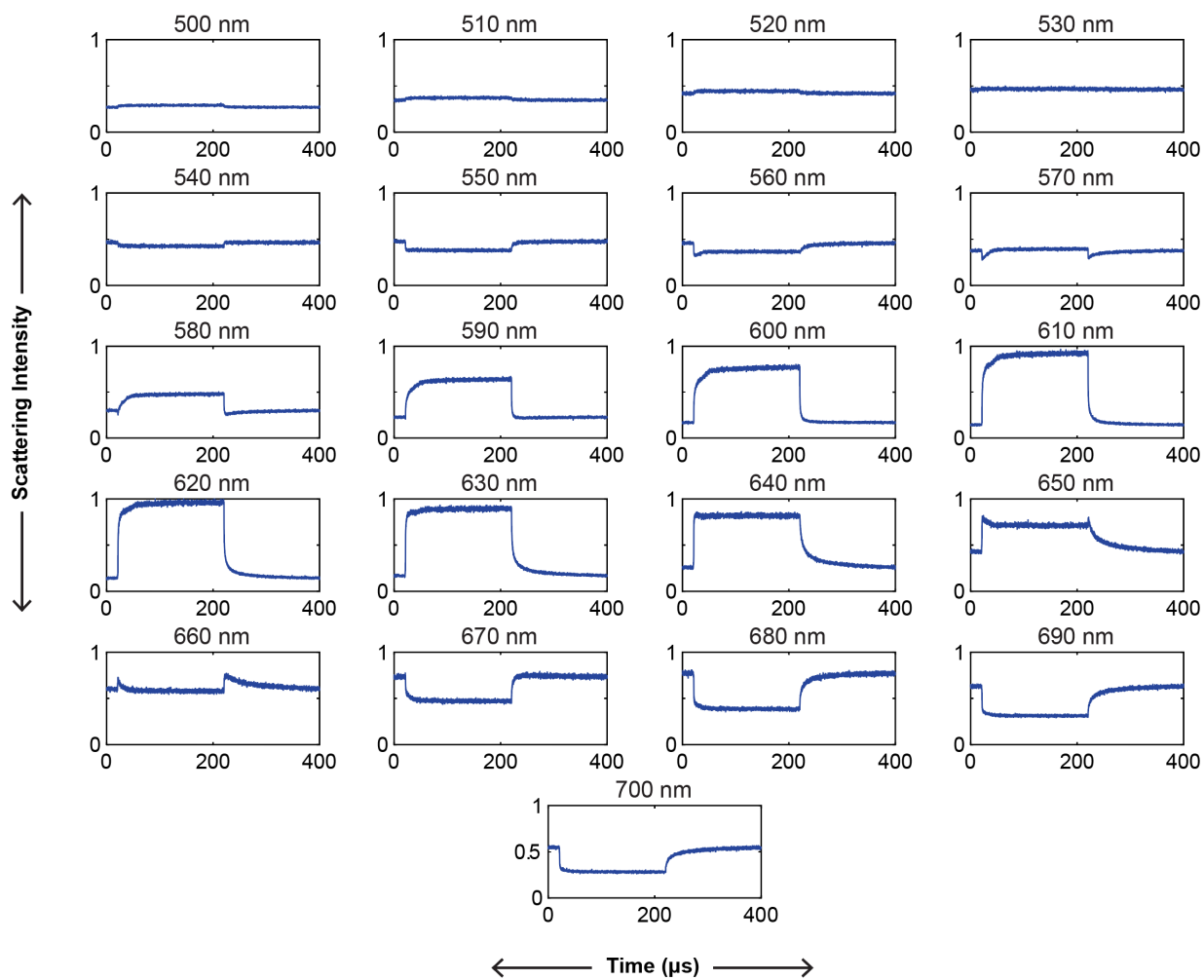

**Fig. S7. Scattering intensity of a single NPoM for various wavelengths during SAW amplitude modulation.**
These signals were acquired by recording the scattered light intensity of a single NPoM as the SAW is switched from off to on to off again. All these traces are taken for the same 400 μs range, but at different illumination wavelengths. For the first ~20 μs, the SAW is off. Then it is turned on for 200 μs, and then off again. This data was used to create Fig. 3A by normalizing each of these traces with the time-averaged spectrum in Fig. 3D and then stitching them together.

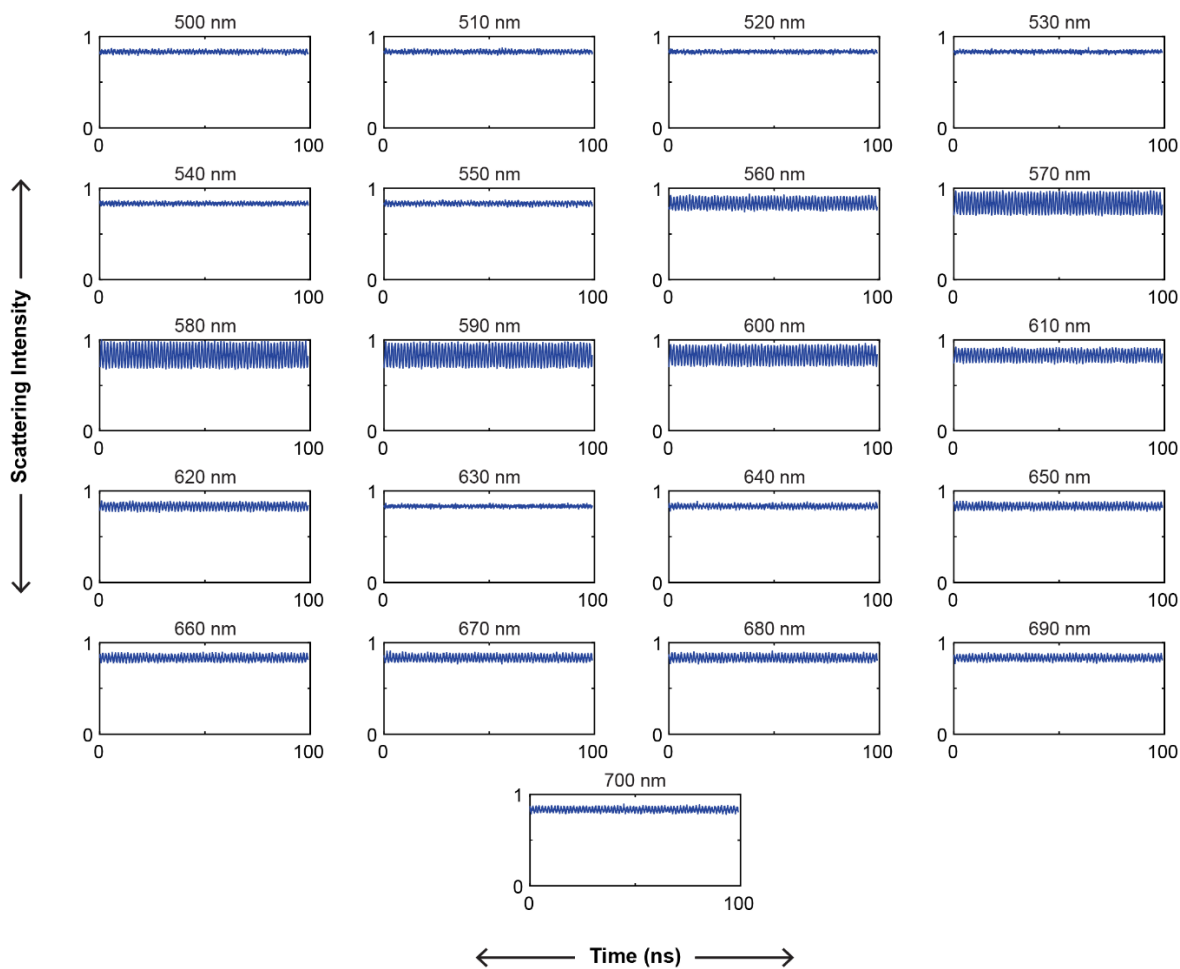

**Fig. S8. Scattering intensity for various wavelengths of a single NPoM during continuous SAW excitation.**
All these traces are taken for the same time range, but at different illumination wavelengths. This data was used to create Fig. 3C by normalizing each of these traces with the time-averaged spectrum in Fig. 3D and then stitching them together.

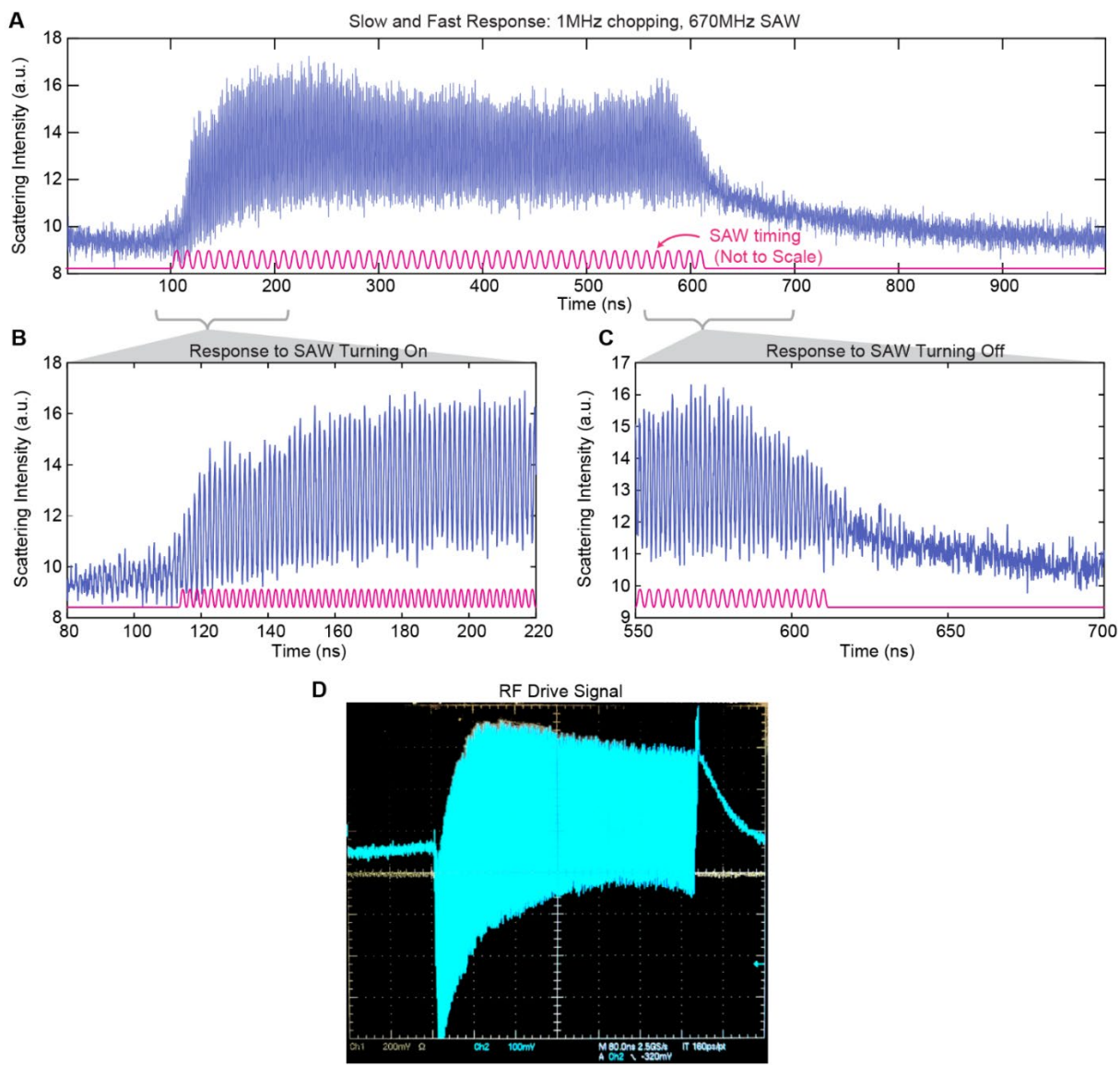

**Fig. S9. Optical scattering at a single wavelength showing both slow and fast optical response.**
(**A**) Scattering of 580 nm light from a single NPoM that is excited with a 670 MHz SAW that is chopped at 1 MHz. The SAW can be switched on or off in about 15 ns. Both the SOR and the FOR are observed here. When the SAW is switched on (**B**), we FOR is immediately seen. As the SAW remains on for about 100 ns, we see the scattering spectrum slowly shifts, which is the SOR. Finally, when the SAW is switched off (**C**), the FOR disappears, and the scattering slowly relaxes to the off scattering state. (**D**) Time trace of the RF signal that is driving the SAW. We can see that the signal is a higher amplitude right after the RF is switched, causing the initial higher amplitude in the FOR observed in (A).

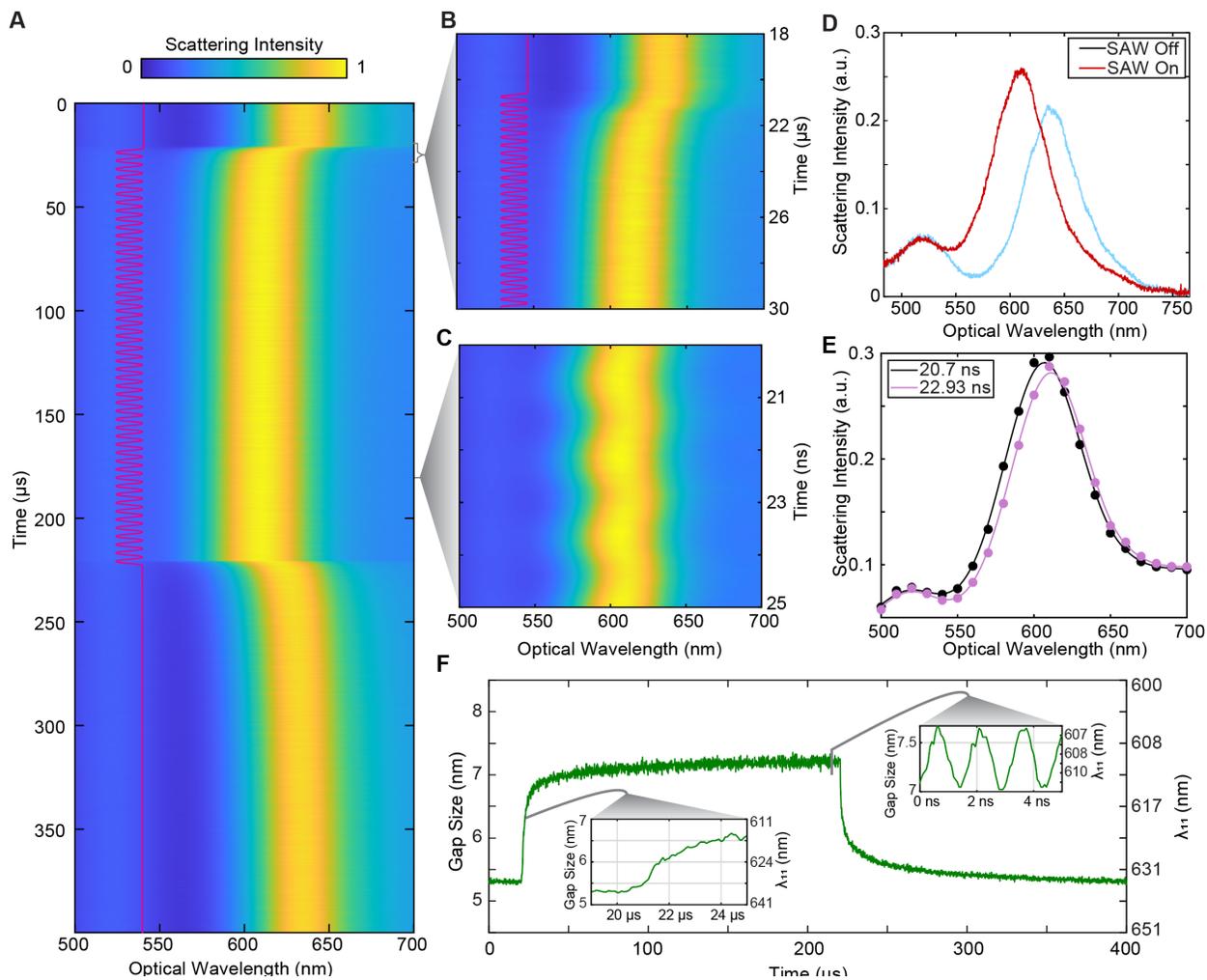

**Fig. S10. Additional time-resolved spectrum for an NPoM with a thicker polymer.**
This is an additional time-resolved spectrum (same format as Fig. 3) for an NPoM fabricated with a thicker polymer layer. The initial polymer thickness is estimated using ellipsometry at 5 to 7 nm thick. The spectrum in the off state is therefore more blue-shifted compared with that of the thinner layer in the NPoM of Fig. 3.

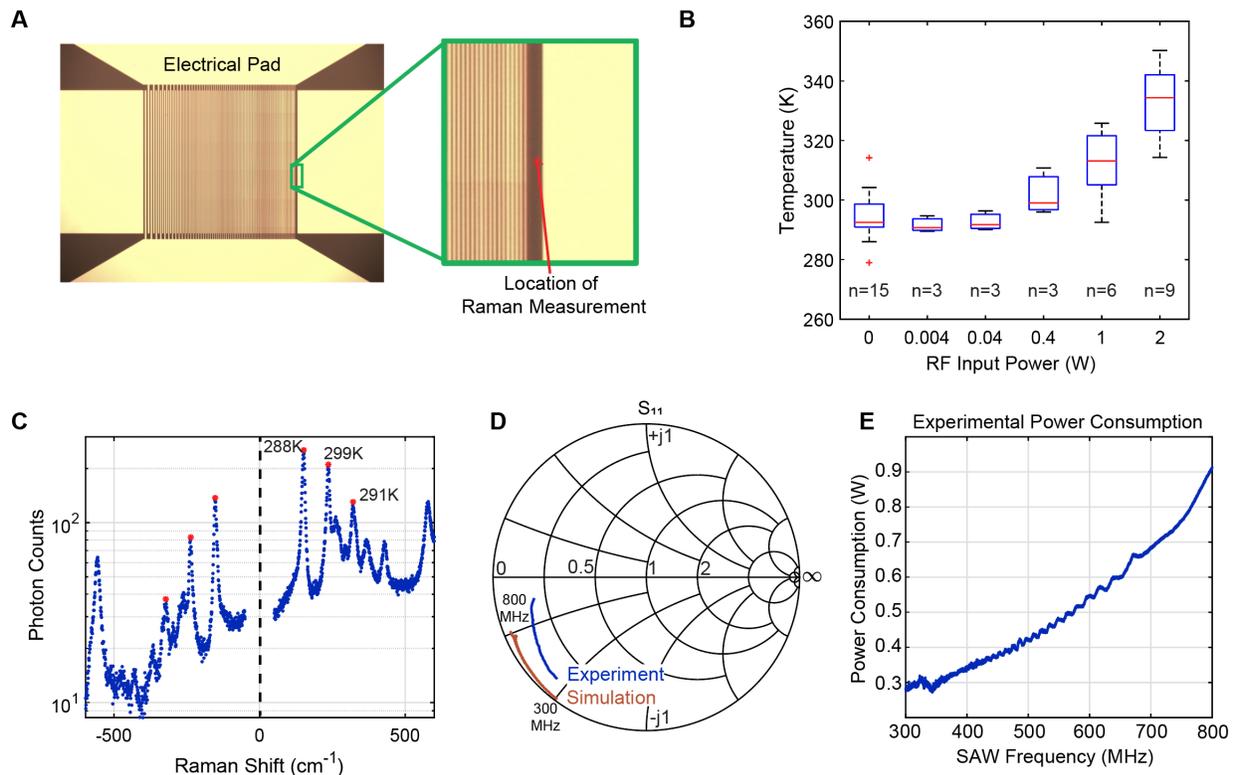

**Fig. S11. Sample surface temperature and power consumption during SAW excitation.**
(**A**) Optical image showing the location of Raman temperature measurement. (**B**) The measured temperature of the device at various RF power levels, all taken with the RF frequency set to 640 MHz. Each boxplot represents the range of temperature calculations for the first three sets of Stokes—anti-Stokes pairs over several measurement repetitions. The total number of data points $n$ in each distribution is shown below the boxplots. (**C**) Example Raman spectrum at room temperature with RF off. The temperatures obtained from three different Stokes—anti-Stokes pairs are displayed next to the Stokes peak in Kelvin. (**D**) RF reflection coefficient $S_{11}$ displayed in a Smith chart for SAW frequencies 300 to 800 MHz. Both simulated and experimental $S_{11}$ are shown. The mount and cabling leading to the device are calibrated out for the experimental spectrum. (**E**) RF power consumption during a typical experiment where 2 W of RF power is sent into the setup. Our devices consume about 0.6 W of RF power to excite a SAW over a 0.5 mm aperture and achieve the largest plasmonic shifts observed.

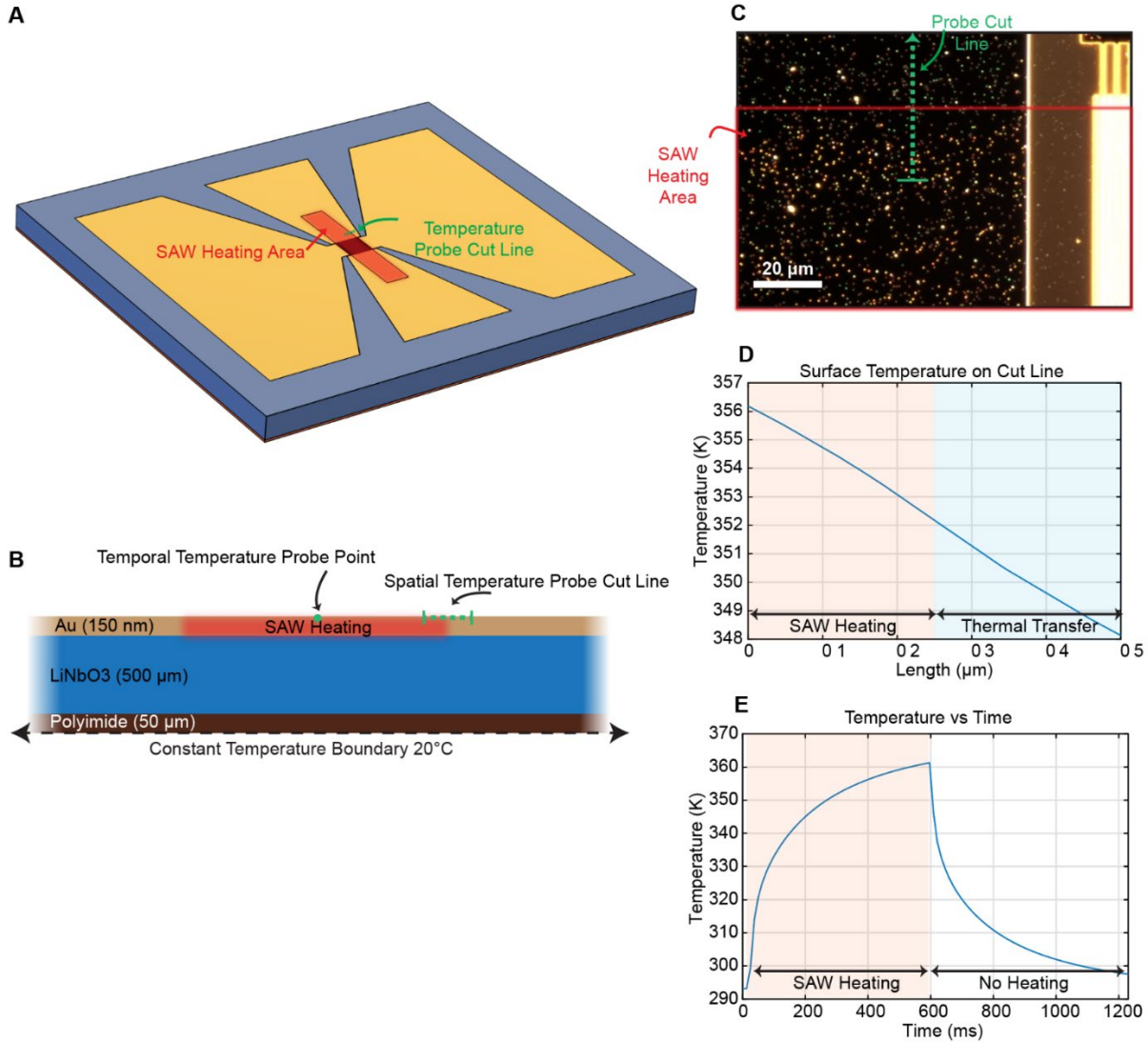

**Fig. S12. Thermal simulations to assess the possible impact of SAW-induced heating on the optical response.**
(**A–B**) Simulation domain that solves the heat equation in solids. SAW/RF heating is modeled as a surface heat source in the metal film that encompasses a $500 \times 3000$ μm area as shown. The temperature is probed at the edge of the heat source as shown as the green dotted line. A 50-μm-thick piece of polyamide tape is on the back of the LiNbO$_3$ substrate. (**C**) Dark-field optical image of the device under SAW activation. The NPoMs in the path of the SAW exhibit a different color than those less than 20 μm away from the path of the SAW. (**D**) Resulting temperature distribution on the cut line at the edge of the SAW heating area. We see that the temperature variations in a 50-μm-sized region near the edge of the SAW beam is less than 10 K. (**E**) Temperature at a point in the middle of the Au film as a function of time immediately after the SAW is turned on. The rise time of the temperature is on the order of 100 ms.

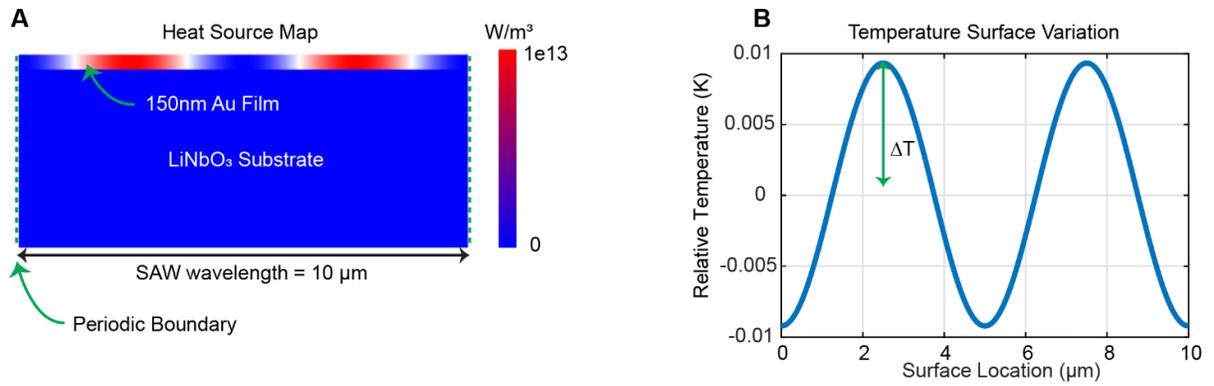

**Fig. S13. Temperature variation on the metal surface generated by a standing acoustic wave.**
(**A**) Simulation domain of a periodic heat source in the Au film that captures the SAW-induced heating from losses in the Au film. (**B**) Resulting temperature distribution caused by the periodic heat source in (A). Due to the high thermal conductivity of the Au film, which disperses the heat, the periodic heat source results in only a minimal temperature difference of 0.1 K across the surface.

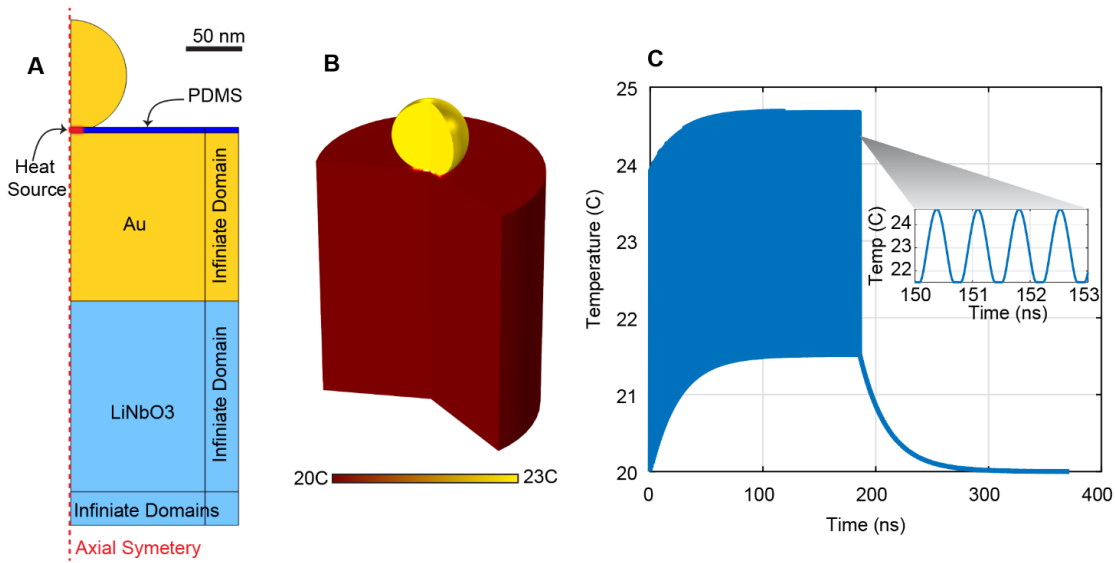

**Fig. S14. Mechanically induced heat generation in the gap of an NPoM.**

(**A**) Simulation domain of an axially symmetric NPoM with a heat source in the gap region that captures the possible heat generation induced by mechanical losses in the gap region (Eq. S28). A basic estimate shows that a maximum of 160 nW is generated in the gap at a SAW frequency of 700 MHz. (**B**) The resulting temperature change due to SAW activation at 700 MHz. (**C**) The maximum temperature of the gap material plotted in time from SAW activation. At $t$ < 0, the SAW is inactive. From time $t$ = 0 to $t$ = 180 ns, the SAW is active, and then switched off again at $t$ = 180 ns. The inset shows the rapidly changing temperature of the gap material as a function of time. The temperature quickly varies at frequency $2f_{saw}$. The slow variation in temperature rises and falls with a decay time constant of around 10 ns. The small temperature variations and their distinct temporal characteristics suggest that heating of the gap region cannot be the cause of the observed color changes in the gap.

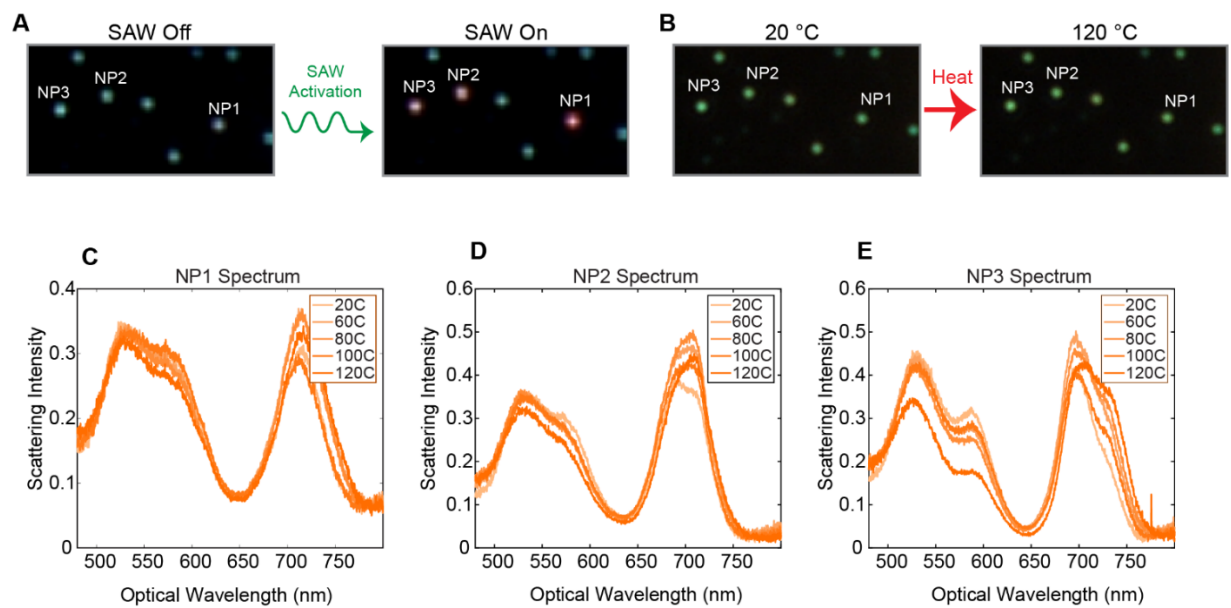

**Fig. S15. Heating of NPoMs on a hot plate and the induced optical responses.**
(**A**) Optical image of three different NPoMs with the SAW inactive and active. These three NPoMs clearly display spectral shifts in the dark-field white light scattering due to the SAW excitation. (**B**) Optical image of the same area before and after heating the device to 120 °C on a hot plate. (**C**) Spectra of the same three NPoMs as in (A) and (B) at different hot plate temperatures. There are no considerable shifts in the spectrum of any NPoM as a result of heating. *Note*: (A) and (B) look subtly different because the images were taken with different camera settings.

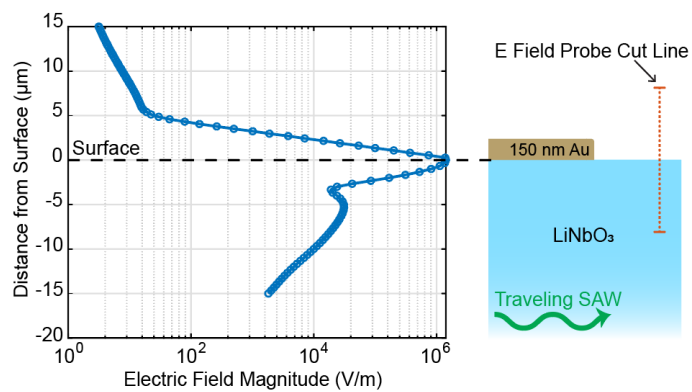

**Fig. S16. Surface electric fields.**
Simulated electric field magnitude in response to 1 W of incident RF power. The electric field profile is shown above the bare LiNbO₃ surface and reaches a maximum of approximately $10^6$ V/m at the surface, which is likely too weak to exert significant electrostatic forces on the AuNPs. Further, in regions covered with the Au film (e.g., where NPoMs are present) the surface electric fields will be screened.

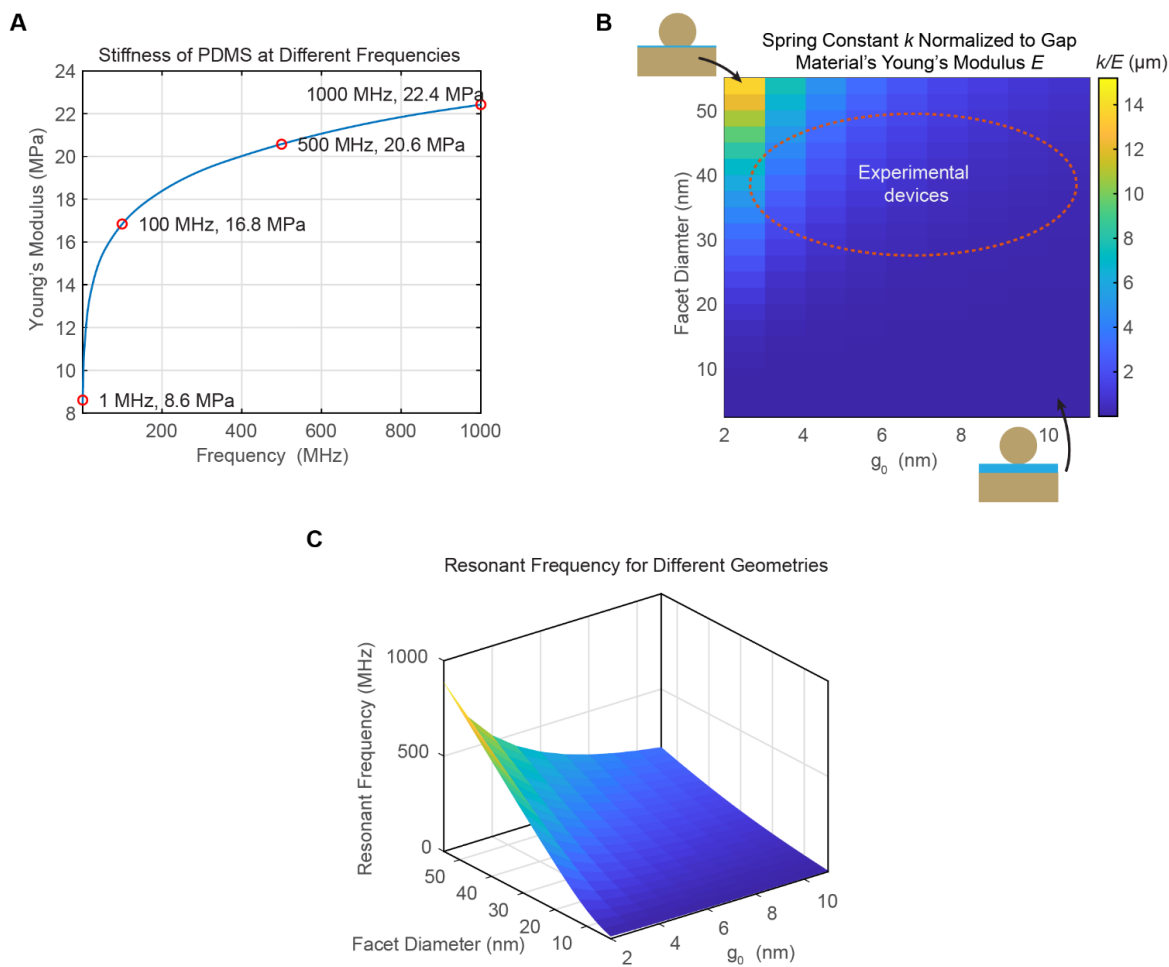

**Fig. S17. Linear elastic properties of PDMS and the NPoM structure.**
(**A**) Reported values of Young's modulus for Sylgard 184 PDMS at MHz frequencies *(37)*. (**B**) The effective spring constant of the polymer layer normalized by the Young's modulus $E$ for a range of different possible NPoM geometries with circular facets. This was obtained using a solid mechanics COMSOL model. (**C**) The resonant frequency of the NPoM structure for $E_{PDMS} = 20.6$ MPa at 500 MHz for a range of different geometries.

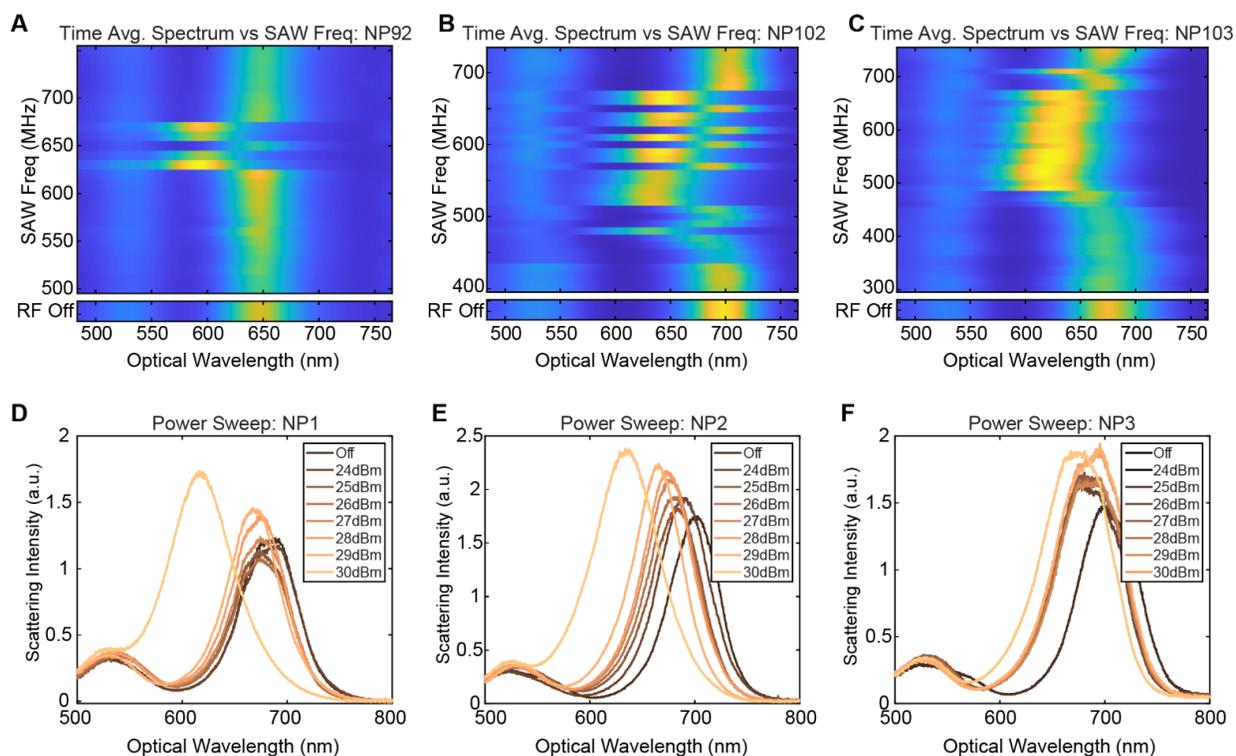

**Fig. S18. Dependence of spectral shifts on SAW power and frequency.**
Time-averaged spectra using unpolarized dark-field illumination from six different NPoMs. Panels (**A-C**) show the spectra at different SAW frequencies, while panels (**D-F**) illustrate the spectra at various SAW excitation powers at a fixed 640 MHz SAW frequency. The data demonstrates that spectral shifts occur across a wide range of frequencies and powers. Additionally, the results reveal a highly nonlinear response. In (D), the nanoparticle does not exhibit a significant spectral shift until the RF power reaches 30 dBm. In contrast, in (F), the spectrum significantly shifts at 24 dBm but remains relatively unchanged from 24 dBm to 29 dBm, with an additional shift occurring only at 30 dBm of excitation. The nonlinearity is further evident in the frequency sweeps (A-C) where slight dips in the emission spectrum of the IDT result in significant reductions in spectral shift.

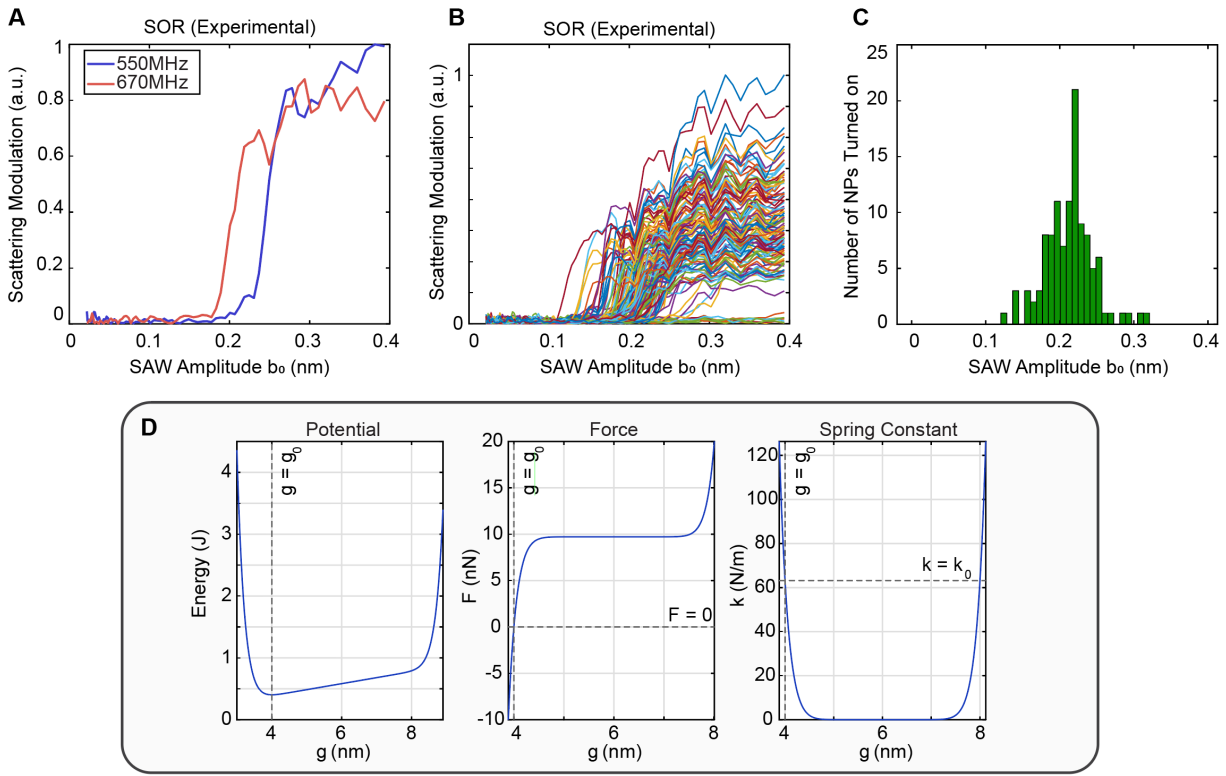

**Fig. S19. Model using a nonlinear spring and asymmetrical dashpot.**
(**A**) SOR for an individual NPoM over a range of SAW amplitudes for two different SAW frequencies. We see a sharp jump up in the SOR around a SAW amplitude of 0.2 nm due to the nonlinear spring. (**B**) Same measurement as in (A) but for 120 random NPoMs on a single device. A total 112 of them demonstrate measurable SOR and they all do so at particular "turn on" amplitudes. (**C**) Histogram of the "turn on" amplitudes for the NPoMs measured in (B). Most of the NPoMs turn on around at an amplitude of 0.2 nm but there is a range of ±0.1 nm. (**D**) Visualization of the force, potential, and spring constant created by a nonlinear spring that can explain the nonlinear dependence of the SOR shown in (A-C). Here, the initial gap size $g_0$ is 4 nm, and the initial spring constant at this gap is $k_0 = 63$ N/m. As the gap increases beyond 4 nm, the spring constant initially decreases then increases again, returning to its initial value again at around 8 nm. On the other hand, as the gap decreases to zero, the NP pushes against the hard substrate itself and thus the spring constant goes to infinity. These results demonstrate that by making both elements of the Voigt model nonlinear, we can closely match the experimental results for individual NPoMs across a range of frequencies and powers.

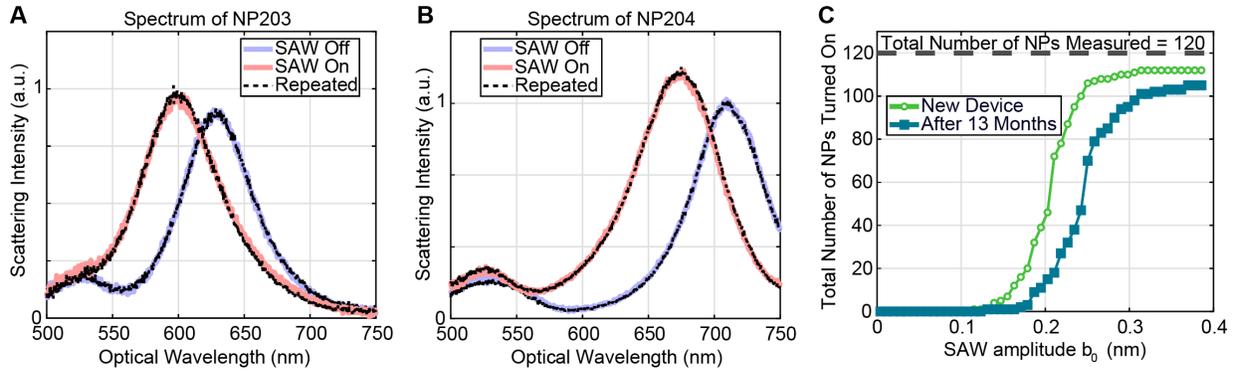

**Fig. S20. Robustness and stability of optical modulation over many cycles.**
(**A**) Time averaged spectrum of a single NPoM when the SAW is switched from off to on at 670 MHz and 2W into the device. The black dotted lines show the repeated spectra after the NPoM underwent 40 billion acoustic cycles. We can see that the optical modulations are robust and do not change significantly. (**B**) Same experiment as in (**A**) but for another nanoparticle. (**C**) The same SAW power sweep performed on the exact same device and set of 120 random NPoMs after 13 months, 5 trillion acoustic cycles, and 10 thousand power cycles. Of the 112 particles initially responsive, 105 remain so. We also observe that slightly more acoustic power is now required to modulate them.

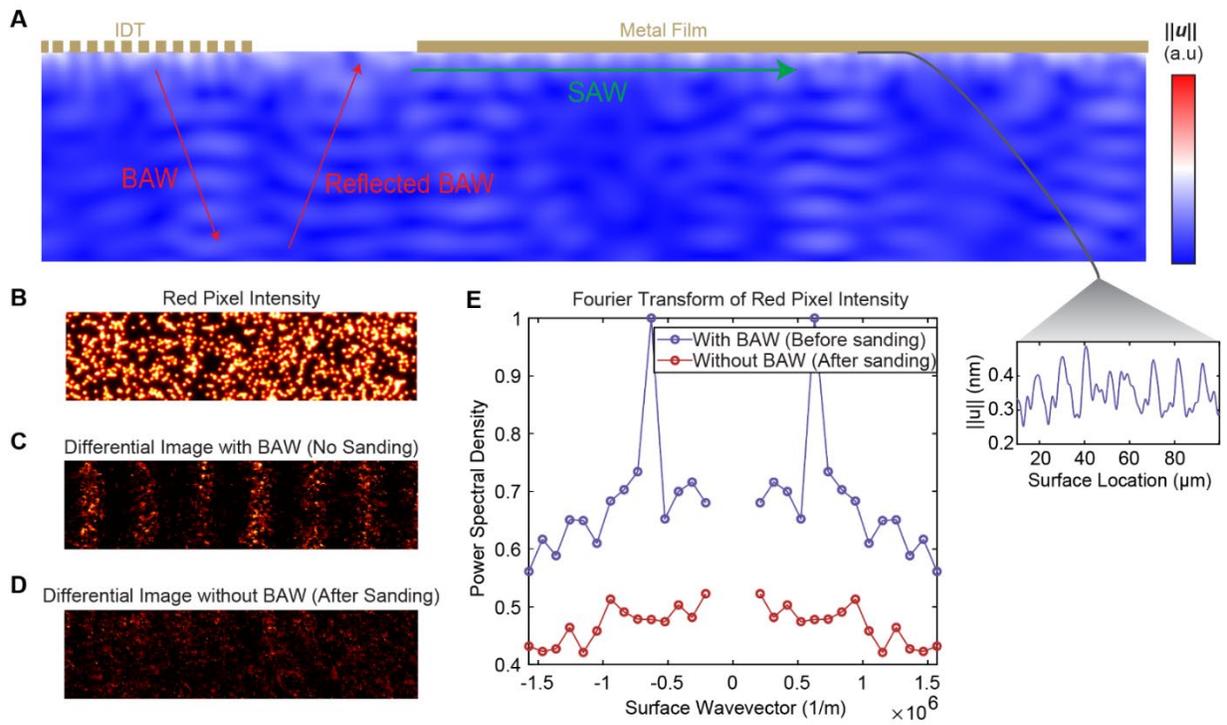

**Fig. S21. Appearance of interference patterns resulting from BAW excitation and their removal.**
(**A**) Simulation similar to that shown in Fig. 2E, but with a free boundary on the bottom of the substrate instead of an acoustically perfectly matched layer. BAWs are reflected from the back of the substrate and propagate back up to the surface, creating an interference pattern with the traveling SAWs. (**B**) Optical image of NPoMs that are activated by SAW at frequency of 500 MHz. (**C**) Difference in the pixel intensity for SAW excitation frequencies of 500 MHz and approximately 510 MHz, showing the appearance of a standing wave pattern. (**D**) Same differential image as shown in (C) but after the back of the substrate was sanded to scatter the BAWs more diffusely. The standing wave pattern is no longer apparent. (**E**) Fourier transform of (C) and (D) showing the standing wave in the frequency domain.

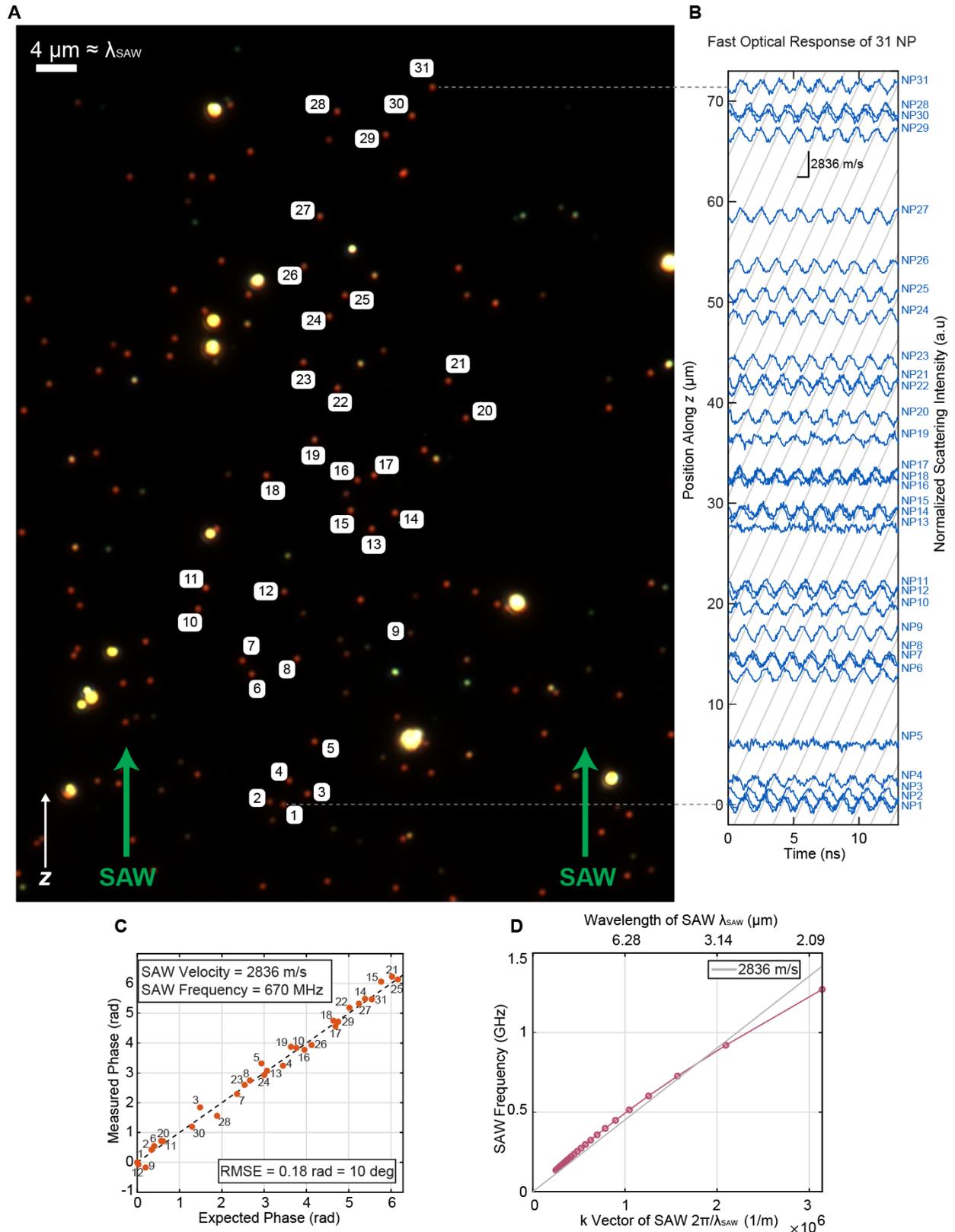

**Fig. S22. Mapping the acoustic phase of a traveling SAW with the fast-optical-response (FOR).**
(**A**) Dark-field photo of 31 NPoMs arranged along the SAW propagation axis (right to left). NPoMs are labeled 1–31; NP31 lies ~ 71 µm upstream of NP1. (**B**) FOR traces for all NPoMs recorded under 610 nm, s-polarized illumination while a 670 MHz SAW is active. The optical modulation phase shifts systematically with position: antennas that are nearly co-located along z (e.g., NP1 and NP2) show almost identical phases, whereas those farther apart (e.g., NP10 and NP11) display larger phase lags. The gray diagonal lines are constant phase contours for a phase velocity of 2836 m s$^{-1}$. These lines intersect with each optical trace at the same phase, therefore implying that the phase velocity across the surface is 2836 m s$^{-1}$. (**C**) Measured optical phase of each NPoM versus the acoustic phase expected from its z-position. A linear fit gives a SAW phase velocity of 2836 m s$^{-1}$ (root-mean-square-error $= 0.18$ rad $\approx 10°$). (**D**) COMSOL simulated SAW dispersion relation for this device (SAW propagating along the z-axis of a Y-cut LiNbO$_3$ substrate coated with 150 nm of Au). The metal layer retards waves with larger in-plane wavevectors because a greater fraction of the mode resides in the metal film. The simulated phase velocity at 670 MHz is 2954 m s$^{-1}$, in reasonable agreement with the experimental value in (C).

**Movie S1. Visualization of acoustically modulated gap plasmon cavities.**

This dark-field optical movie illustrates the device's response to activation by SAWs. The IDT and the region subjected to the collimated SAW beam are highlighted in green and blue, respectively. The SAW is switched on and off at 1 Hz, while its frequency is continuously swept between approximately 585 MHz and 610 MHz. Upon SAW activation, a discernible alteration in light scattering is observed at the NPoM sites within the SAW beam's influence. In contrast, NPoMs located outside the SAW beam exhibit no color change, confirming that the observed scattering modulation is directly attributable to the presence of the SAW. The thickness of the polymer layer on this sample is measured to be approximately 4 to 7 nm.

**Movie S2. Visualization of collimated SAW.**

Video of device shown in Fig. S12C. The SAW power is chopped at 1 Hz and the frequency is swept from approximately 650 MHz to 680 MHz over a 10 s time window. NPoMs exposed to the SAW change color with SAW activation, while NPoMs out of the SAW beam are unaffected.

**Movie S3. SAW and BAW interference producing a visible standing wave pattern.**

Dark-field movie of a device activated by SAWs entering from the right and propagating to the left. The SAW frequency sweeps continuously and repeatedly from 450 MHz to 550 MHz over a 2.5-second interval. Dynamic standing wave patterns in optical scattering appear and shift in response to the changing SAW frequency, likely due to interference between the incident SAW and BAWs reflected from the substrate's rear surface. Sanding the back of the substrate with 240-grit sandpaper scatters the BAWs and eliminates these standing wave patterns (see Fig. S21).